\documentclass[a4paper,12pt]{article}

\usepackage{amsmath,graphicx,cite}

\allowdisplaybreaks[1]

\title{
\vspace{-1.5cm}
\begin{flushright}
{\small ROMA-1447-06}
\end{flushright}
\vspace{0.5cm}
\bf Neutrino flux ratios at neutrino telescopes: The
role of uncertainties of neutrino mixing parameters and applications
to neutrino decay}

\author{\large{\sc Davide~Meloni${}^{1,}$\footnote{Email: {\tt
	meloni@roma1.infn.it}}~~and
	Tommy~Ohlsson${}^{2,}$\footnote{Email: {\tt
	tommy@theophys.kth.se}}}}

\date{}

\begin{document}

\maketitle

{\it
\footnotemark[1]%
INFN, Sezione di Roma and Dipartimento di Fisica, Universit{\`a} ``La
Sapienza'',\\
IT-00185~Rome, Italy

\footnotemark[2]%
Department of Theoretical Physics, School of
Engineering Sciences,\\
Royal Institute of Technology (KTH) -- AlbaNova
University Center,\\
Roslagstullsbacken 21, SE-106~91~Stockholm, Sweden
}

\abstract
\noindent
In this paper, we derive simple and general perturbative formulas for
the flavor flux ratios
$R_{\alpha\beta}=\phi_{\nu_\alpha}/\phi_{\nu_\beta}$ that could be
measured at neutrino telescopes. We discuss in detail the role of the
uncertainties of the neutrino mixing parameters showing that they have
to be seriously taken into account in any realistic discussion about
flavor measurements at neutrino telescopes. In addition, we analyze
the impact of such uncertainties in telling the standard neutrino
oscillation framework from scenarios involving, {\it e.g.}, neutrino
decay and we find that the ratio $R_{e\mu}$ is the most sensitive one
to ``new physics'' effects beyond the Standard Model. We also compute
the more realistic muon-to-shower ratio for a particular configuration
of the IceCube experiment, observing that using this experimental
quantity a clear separation between standard and non-standard neutrino
physics cannot be obtained.

\section{Introduction}

The standard framework of neutrino oscillations is a successful
description of data from recent neutrino experiments. However, there
might be subleading effects that are not covered by neutrino
oscillations, since the experimental data are still impaired by rather
large uncertainties. Such subleading effects could be described by
introducing an extended neutrino oscillation framework including
so-called damping effects that could stem from, {\it e.g.}, neutrino
decay or neutrino decoherence. In this work, we will introduce damping
factors, describing the damping effects, in a phenomenological way as
additional factors to the ordinary terms in the formulas for the
neutrino oscillation probabilities. A promising situation to look for
such subleading effects would be from neutrino sources at large
distances, since in this case the result of the damping factors will
be most visible because of averaging. In addition, it could be an
alternative way to measure the fundamental neutrino parameters, but
not the neutrino mass-squared difference, since the averaged neutrino
oscillation probabilities do not depend on them. Therefore, we will
investigate neutrino oscillations with damping effects in general, and
different scenarios in which damping effects arise from neutrino decay
in particular. We will study these scenarios in the case of neutrinos
coming from large distances. The most plausible way to measure such
neutrinos would be at neutrino telescopes in ice
\cite{Achterberg:2006md} or water \cite{ Katz:2006wv}. Many papers in
the literature have been devoted to study the dependence of the
neutrino fluxes on the neutrino mixing parameters, namely the three
leptonic mixing angles $\theta_{12}$, $\theta_{13}$, and $\theta_{23}$
as well as the CP-violating phase $\delta$ (for an incomplete list of
references, see
Refs.~\cite{Athar:2000yw,Farzan:2002ct,Bhattacharjee:2005nh,Serpico:2005sz,Serpico:2005bs,Kachelriess:2006fi,Rodejohann:2006qq}).
The usual starting point is that neutrinos are produced via decays of
pions and kaons created by hadronic ({\it i.e.}, $pp$ collisions) and
photohadronic ($p\gamma$) interactions \cite{Ginzburg:1990sk}, leading
to the well-known flux ratios at the source
$\phi^0_{\nu_e}:\phi^0_{\nu_\mu}:\phi^0_{\nu_\tau}=1:2:0$. However,
other effects, like muon energy loss in strong magnetic fields
\cite{Rachen:1998fd,Kashti:2005qa} can noticeably alter the flavor
composition; in addition, a different flux ratio at the source can be
obtained for the so-called ``neutron beam sources'' in which neutrinos
are produced from neutron decays
\cite{Anchordoqui:2003vc,Hooper:2004xr}. In any case, on their way
from the source to the Earth, neutrinos oscillate and the fluxes
arriving at the detector acquire a dependence on the neutrino mixing
parameters which can, in principle, be used to improve our knowledge
on the fundamental neutrino parameters and/or used in connection with
reactor experiments and neutrino beams to better the determination of
$\delta$ and the neutrino mass hierarchy. However, it could happen at
the time when neutrino telescopes will be operational that the
uncertainties of the neutrino mixing parameters could be large enough
not to be neglected in theoretical estimates of the arriving fluxes
\cite{Winter:2006ce}. This implies that any possible dependence on the
neutrino mixing parameters can be completely hidden, and what is also
important, that any {\it small} deviation from the expected number of
neutrinos can simply be the effect of our ignorance about the precise
values of the neutrino mixing parameters and not due to any ``new
physics'' effects in large distance neutrino oscillations, as those
given, {\it e.g.}, by neutrino decay
\cite{Acker:1991ej,Beacom:2002vi}, breakdown of fundamental symmetries
\cite{Hooper:2004xr,Hooper:2005jp,Gonzalez-Garcia:2005xw,Anchordoqui:2005gj},
or pseudo-Dirac nature of neutrinos
\cite{Kobayashi:2000md,Beacom:2003eu}.

For this reason, in this work, we analyze in some detail the impact of
the current neutrino mixing parameter errors on the determination of
the fluxes arriving at neutrino telescopes. We explicitly illustrate
the dependence on the deviation from the best-fit values of
$\theta_{12},\theta_{13}$, and $\theta_{23}$ up to second order, in
the case that the fluxes at the source are in the ratios
$\phi^0_{\nu_e}:\phi^0_{\nu_\mu}:\phi^0_{\nu_\tau}=1:2:0$. For
comparison, we also show how the fluxes on Earth change if neutrino
decay takes place and we discuss whether it is possible to distinguish
them from the fluxes computed in the standard neutrino oscillation
framework, and to what extent.

Our paper is organized as follows: In Sec.~\ref{sec:damping}, we
derive the averaged neutrino oscillation probabilities including
damping factors. Next, in Sec.~\ref{sec:flux}, we analyze the
three-flavor flux ratios and study the impact of the uncertainties of
the fundamental neutrino parameters. Then, in Sec.~\ref{sec:decay}, we
investigate the flux ratios with neutrino decay, present the different
neutrino decay scenarios, as well as we discuss the possibility to
detect the subleading effects at neutrino telescopes. Finally, in
Sec.~\ref{sec:summary}, we summarize our work as well as we present
our conclusions.

\section{Derivation of averaged neutrino oscillation probabilities
including damping factors}
\label{sec:damping}

In general, the neutrino oscillation probabilities in the standard
three-flavor framework have the following form:
\begin{equation}
P_{\alpha\beta} = \sum_{i=1}^3 \sum_{i=j}^3 J_{\alpha\beta}^{ij}
\exp(-{\rm i} \Phi_{ij}),
\label{eq:Pab}
\end{equation}
where $J_{\alpha\beta}^{ij} = U_{\alpha j} U_{\beta j}^\ast U_{\alpha
i}^\ast U_{\beta i}$ and $\Phi_{ij} = \Delta m_{ij}^2 L/(2E)$. Here
$U$ is the leptonic mixing matrix, $\Delta m_{ij}^2 = m_i^2 - m_j^2$
are the neutrino mass-squared differences, $L$ is the baseline length,
and $E$ is the neutrino energy. The standard parameterization of the
leptonic mixing matrix is given by
\begin{equation}
U = \left(
  \begin{matrix} c_{12} c_{13} & s_{12} c_{13} & s_{13} {\rm e}^{-{\rm
  i} \delta_{\rm CP}} \\ -s_{12} c_{23} - c_{12} s_{13} s_{23} {\rm
  e}^{{\rm i} \delta_{\rm CP}} & c_{12} c_{23} - s_{12} s_{13} s_{23}
  {\rm e}^{{\rm i} \delta_{\rm CP}} & c_{13} s_{23} \\ s_{12} s_{23} -
  c_{12} s_{13} c_{23} {\rm e}^{{\rm i} \delta_{\rm CP}} & -c_{12}
  s_{23} - s_{12} s_{13} c_{23} {\rm e}^{{\rm i} \delta_{\rm CP}} &
  c_{13} c_{23} \end{matrix} \right),
\end{equation}
where $c_{ij}\equiv \cos \theta_{ij}$, $s_{ij} \equiv \sin \theta_{ij}$, and
$\delta_{\rm CP}$ is the Dirac CP-violating phase.
In the case that the ratio $L/E$ is
large, {\it i.e.}, $L/E \gg 2(\Delta m_{ij}^2)^{-1}$, one obtains in
vacuum\footnote{Matter effects inside the source can affect the
neutrino transition probabilities, see Ref.~\cite{Mena:2006eq}.} the
following averaged neutrino oscillation probabilities
\begin{equation}
\langle P_{\alpha\beta} \rangle = \sum_{i=1}^3 J_{\alpha\beta}^{ii} =
\sum_{i=1}^3 |U_{\alpha i}|^2 |U_{\beta i}|^2.
\label{eq:<Pab>}
\end{equation}
Note that these probabilities depend only on the parameters
$J_{\alpha\beta}^{ij}$, where $i = j$, as well as they are independent
of the oscillation frequencies, since these have been averaged out to
zero for large values of the ratio $L/E$.

Now, introducing damping factors $D_{ij}$ in Eq.~(\ref{eq:Pab}), the
neutrino oscillation probabilities can be written as
\begin{align}
P_{\alpha\beta} &= \sum_{i=1}^3 \sum_{j=1}^3 D_{ij}
J_{\alpha\beta}^{ij} \exp(-{\rm i} \Phi_{ij}) \nonumber\\
&= \sum_{i=1}^3 D_{ii} J_{\alpha\beta}^{ii} + 2 \sum_{1 \leq i,j \leq
3} D_{ij} |J_{\alpha\beta}^{ij}| \cos(\Phi_{ij} + \arg
J_{\alpha\beta}^{ij}),
\end{align}
where the damping factors are given by
\begin{equation}
D_{ij} = \exp\left( - \alpha_{ij} \frac{|\Delta m_{ij}^2|^\xi
L^\beta}{E^\gamma} \right)
\end{equation}
with $\alpha_{ij}$ being elements in a non-negative damping
coefficient matrix. We will assume $\alpha_{ij}$ to be independent of
energy. Note that both the oscillation frequencies $\Phi_{ij}$ and the
damping factors $D_{ij}$ are functions of the baseline length $L$ and
the neutrino energy $E$. The oscillation frequencies are functions of
the ratio $L/E$ only, whereas the damping factors are more general
functions of the parameters $L$ and $E$, {\it i.e.}, $D_{ij} =
D_{ij}(L^\beta/E^\gamma)$.

Similarly, as in the averaging procedure above for the standard
framework, one can perform an averaging of the neutrino oscillation
probabilities including damping factors for large values of the ratio
$L/E$, {\it i.e.}, $L/E \gg 2 (\Delta m_{ij}^2)^{-1}$. Note that it
would be unnatural to include damping effects after one has performed
averaging. Thus, in the case when $\beta = \gamma$, the damping
factors are functions of $(L/E)^\beta$ only, and therefore, the
following averaging (with $\ell = L/E$) can be
introduced\footnote{Note that averaging with respect to large
distances can be obtained by simply assuming $E$ to be a constant,
{\it i.e.}, by defining $\ell = \frac{1}{E} L$, where $E$ is a
constant.}
\begin{equation}
\langle P_{\alpha\beta} \rangle = \lim_{x \to \infty} \frac{\int_0^x
P_{\alpha\beta}(\ell) \, {\rm d}\ell}{\int_0^x {\rm d}\ell} = \lim_{x
\to \infty} \frac{1}{x} \int_0^x P_{\alpha\beta}(\ell) \, {\rm d}\ell.
\end{equation}
In this work, we will not consider the case when $\beta \neq
\gamma$. Partly, because this case has less obvious applications. In
general, for an arbitrary value of the parameter $\beta$, it is not
possible to compute the averaged neutrino oscillation probabilities
including damping factors, but it is possible for the cases when $\beta =
1$ and $\beta = 2$, which are the cases that are important for
applications to scenarios that could arise in Nature, see
Ref.~\cite{Blennow:2005yk}.

In the case when $\beta = 1$, we obtain
\begin{equation}
\langle P_{\alpha\beta} \rangle = \lim_{x \to \infty} \sum_{i=1}^3
\sum_{j=1}^3 \langle D_{ij} \rangle (x) J_{\alpha\beta}^{ij},
\label{eq:Pab2}
\end{equation}
where the effective damping factors are given by
\begin{equation}
\langle D_{ij} \rangle (x) = \frac{1 - \exp\left(-{\rm i} \frac{\Delta
m_{ij}^2}{2}x- \alpha_{ij} |\Delta m_{ij}^2|^\xi x\right)}{{\rm i}
\frac{\Delta m_{ij}^2}{2}x+ \alpha_{ij} |\Delta m_{ij}^2|^\xi x},
\end{equation}
which can be written in a more simpler form as
\begin{equation}
\langle D_{ij} \rangle (x) = \sum_{n=0}^\infty \frac{1}{n+1}
\frac{(-a_{ij} x)^n}{n!},
\end{equation}
where
\begin{equation}
a_{ij} = {\rm i} \frac{\Delta m_{ij}^2}{2}+ \alpha_{ij} |\Delta m_{ij}^2|^\xi.
\end{equation}
For $\xi \neq 0$, the parameters $a_{ij}$ are in general non-zero if
$i \neq j$, and therefore, one finds that $\lim_{x \to \infty} \langle
D_{ij} \rangle (x) = 0$, whereas $a_{ij} = 0$ if $i = j$, which means
that $\langle D_{ii} \rangle (x) = 1$, and hence, one regains the
formula in Eq.~(\ref{eq:<Pab>}). On the other hand, for $\xi = 0$,
$a_{ij} = \alpha_{ij} + {\rm i} \frac{\Delta m_{ij}^2}{2}$ if $i \neq
j$, and especially, $a_{ij} = \alpha_{ij}$ if $i = j$. Thus, in the
case that $\xi = 0$ and in the limit $x \to \infty$, we find that
$\langle P_{\alpha\beta} \rangle = 0$, since $\lim_{x \to \infty}
\langle D_{ij} \rangle (x) = 0$ for all $x$ if $i = j$ and $\lim_{x
\to \infty} \langle D_{ij} \rangle (x) = 0$ if $i \neq j$ and
$\alpha_{ij} \neq 0$. However, if $\alpha_{ij} = 0$ for fixed $i = j$,
then $\lim_{x \to \infty} \langle D_{ii} \rangle (x) = 1$. Thus, in
the case that $\xi = 0$ and $\alpha_{ii} = 0$ for fixed $i$, the
averaged neutrino oscillation probabilities become
\begin{equation}
\langle P_{\alpha\beta} \rangle = J_{\alpha\beta}^{ii} = |U_{\alpha
i}|^2 |U_{\beta i}|^2.
\end{equation}
In general, if $\alpha_{ii} = 0$ for more than one fixed $i$, then one obtains
\begin{equation}
\langle P_{\alpha\beta} \rangle = \sum_{i \in N} J_{\alpha\beta}^{ii}
= \sum_{i \in N} |U_{\alpha i}|^2 |U_{\beta i}|^2,
\label{eq:Pab3}
\end{equation}
where $N$ can be any of the sets $\{1,2\}$, $\{1,3\}$, $\{2,3\}$, and
$\{1,2,3\}$. In the case when $N = \{1,2,3\}$, one recovers the
standard framework formula in Eq.~(\ref{eq:<Pab>}).

Similarly, in the case when $\beta = 2$, we can again write the formula for
the averaged neutrino oscillation probabilities including damping factors
as in Eq.~(\ref{eq:Pab2}), but now the effective damping factors are given by
\begin{align}
\langle D_{ij} \rangle (x) &= \frac{1}{2 k_{ij} x} \sqrt{\pi}
\exp\left[- \frac{(\Delta m_{ij}^2)^2}{16 \alpha_{ij} k_{ij}^2}\right]
\nonumber\\
&\times \left[ {\rm erf}\left( k_{ij} x + {\rm i} \frac{\Delta
m_{ij}^2}{4 k_{ij}} \right) - {\rm i} \, {\rm erfi}\left(
\frac{\Delta m_{ij}^2}{4 k_{ij}} \right) \right],
\end{align}
where $k_{ij} = \sqrt{\alpha_{ij} |\Delta
m_{ij}^2|^\xi}$. Note that the argument of the imaginary
error function is independent of $x$. Again, the situation is the same
as in the case when $\beta = 1$, which means that non-zero averaged
neutrino oscillation probabilities can only be obtained for $\xi = 0$
and $\alpha_{ii} = 0$ for fixed $i$.

\section{Neutrino flux ratios}
\label{sec:flux}

In the cases $\beta = \gamma = 1$ and $\beta = \gamma = 2$, the
averaged neutrino oscillation probabilities with damping factors
included that are described in Eq.~(\ref{eq:Pab3}) can effectively be
written as
\begin{equation}
\langle P_{\alpha\beta} \rangle = \sum_{i=1}^3 d_{i} J_{\alpha\beta}^{ii},
\label{eq:<Pab>eff}
\end{equation}
where $d_{i}$ are the normalized damping factors, which can have
the value 0 or 1. Of course, if $d_i = 1$ for $i=1,2,3$, then one
finds the averaged neutrino oscillation probabilities in the standard
framework, which are given in Eq.~(\ref{eq:<Pab>}). 

Starting from the flux ratios at the source
$\phi^0_{\nu_e}:\phi^0_{\nu_\mu}:\phi^0_{\nu_\tau}=1:2:0$, the
neutrino fluxes arriving at a detector are sensitive to neutrino
oscillation in vacuum and are then computed as
\begin{eqnarray}
\phi_{\nu_e} & =&  \langle P_{ee}\rangle + 2 \,
\langle P_{\mu e} \rangle, \nonumber \\
\label{fluxesold}
\phi_{\nu_\mu} & =& \langle P_{e\mu}\rangle + 2 \,\langle
P_{\mu\mu}\rangle, \\
\phi_{\nu_\tau} &=&
\langle P_{e\tau}\rangle + 2 \,\langle P_{\mu\tau}\rangle. \nonumber
\end{eqnarray}
Then, the three-flavor flux ratios analyzed in this work 
are defined as follows:
\begin{equation}
\label{ratios}
R_{e\mu} = \frac{\phi_{\nu_e}}{\phi_{\nu_\mu}}, \quad R_{e\tau} =
\frac{\phi_{\nu_e}}{\phi_{\nu_\tau}}, \quad R_{\mu\tau} =
\frac{\phi_{\nu_\mu}}{\phi_{\nu_\tau}}.
\end{equation}
Although the flux ratios are not independent of each other, since a
measurement of two of them will give the value of the third, we prefer
to discuss them separately.

\subsection{Standard neutrino framework}

In this subsection, we analyze the analytical form of three-flavor
flux ratios on Earth. These ratios depend on three mixing angles and
one CP-violating phase, as implied by Eq.~(\ref{fluxesold}), and are,
in principle, sensitive to subleading effects (encoded in the
dependence of $\theta_{13}$) and the errors of the mixing angles
$\theta_{12}$ and $\theta_{23}$. For maximal mixing of $\theta_{23}$,
{\it i.e.}, $\theta_{23}=\pi/4$, and vanishing $\theta_{13}$, the
robust prediction of Eq.~(\ref{fluxesold}) is
$\phi_{\nu_e}:\phi_{\nu_\mu}:\phi_{\nu_\tau}=1:1:1$
\cite{Athar:2000yw}, independently of $\theta_{12}$. Going beyond this
{\it zero-order} approximation allows us not only to study the role of
non-maximal $\theta_{23}$ and non-vanishing $\theta_{13}$, but also to
understand the impact of introducing the uncertainties of the neutrino
mixing parameters.

\begin{figure}
\centerline{\includegraphics[width=0.65\textwidth]{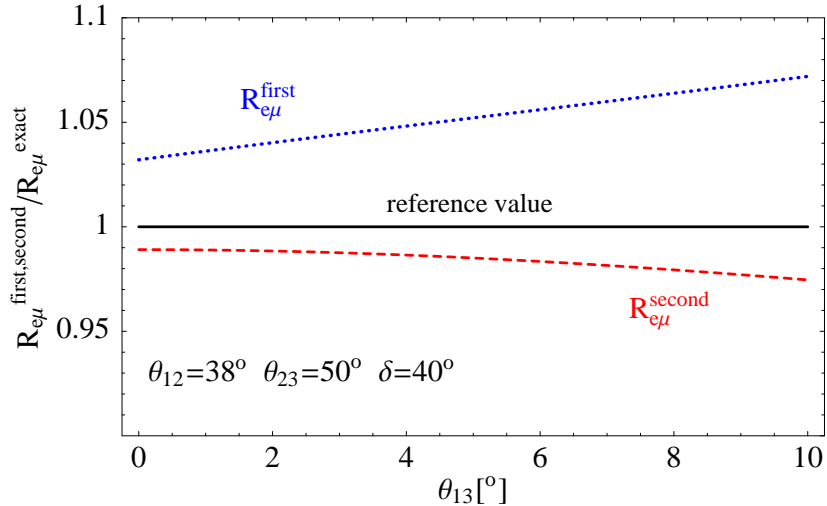}}
\caption{\it Comparison between the approximate formulas $R^{\rm
first}_{e\mu}$ and $R^{\rm second}_{e\mu}$ as well as the exact
formula for $R_{e\mu}$. The solid curve corresponds to the reference
value 1, whereas the dotted and dashed curves correspond to the first-
and second-order approximations, respectively.}
\label{comparison}
\end{figure}

Since the exact formulas for the flux ratios are quite cumbersome, we
prefer to present our results in some useful approximations. {}From
experimental data, we know that $\theta_{13}$ is small and
$\theta_{23}$ is close to maximal mixing. For that reason, we can
expand the ratios in $\theta_{13}$ and
$\delta_{23}=\theta_{23}-\pi/4$, parametrizing the deviation from
maximal 23-mixing \cite{Xing:2006xd}. At the same time, we can also
expand for small $\delta_{12}=\theta_{12}-\bar \theta_{12}$, $\bar
\theta_{12}$ being the best-fit value for $\theta_{12}$. No
restrictions have been applied on the CP-violating phase $\delta$,
which means that the following formulas are valid to all orders in
$\delta$. Up to second order in the small quantities $\theta_{13}$,
$\delta_{12}$, and $\delta_{23}$, they read
\begin{align}
R_{e\mu} &= 1+\frac{3}{4} \cos (\delta ) \sin \left(4
\theta_{12}\right) \theta_{13}-\frac{3}{2} \sin^2\left(2
\theta_{12}\right) \delta_{23}\nonumber\\
&+\frac{1}{8} \cos^2(\delta) \left[3 \cos \left(4
\theta_{12}\right)-5\right] \sin^2\left(2 \theta_{12}\right)
\theta_{13}^2\nonumber\\
&+\frac{1}{32} \left[-28 \cos \left(4 \theta_{12}\right)+3 \cos
\left(8 \theta_{12}\right)-103\right] \delta_{23}^2\nonumber\\
&+3 \cos (\delta ) \cos \left(4 \theta_{12}\right) \delta_{12}
\theta_{13}-3 \sin \left(4 \theta_{12}\right) \delta_{12}
\delta_{23}\nonumber\\
&+\frac{1}{16} \cos (\delta) \left[3 \sin \left(8
\theta_{12}\right)-22 \sin \left(4 \theta_{12}\right)\right]
\theta_{13} \delta_{23}, \label{emuexp}\\
R_{e\tau} &= 1+\frac{3}{4} \cos (\delta ) \sin \left(4
\theta_{12}\right) \theta_{13}-\frac{3}{2} \sin^2\left(2
\theta_{12}\right) \delta_{23}\nonumber\\
&+\frac{1}{8} \cos^2(\delta) \left[3 \cos \left(4
\theta_{12}\right)+11\right] \sin^2\left(2 \theta_{12}\right)
\theta_{13}^2\nonumber\\
&+\frac{1}{32} \left[4 \cos \left(4 \theta_{12}\right)+3 \cos \left(8
\theta_{12}\right)+121\right] \delta_{23}^2\nonumber\\
&+3 \cos (\delta ) \cos \left(4 \theta_{12}\right) \delta_{12}
\theta_{13}-3 \sin \left(4 \theta_{12}\right) \delta_{12}
\delta_{23}\nonumber\\
&+\frac{1}{16} \cos (\delta) \left[10 \sin \left(4
\theta_{12}\right)+3 \sin \left(8 \theta_{12}\right)\right]
\theta_{13} \delta_{23}, \label{etauexp}\\
R_{\mu\tau} &= 1\nonumber\\
&+2 \cos^2(\delta )\sin^2\left(2 \theta_{12}\right)
\theta_{13}^2+\left[\cos \left(4 \theta_{12}\right)+7\right]
\delta_{23}^2\nonumber\\
&+2 \cos (\delta ) \sin \left(4 \theta_{12}\right) \theta_{13} \delta_{23}.
\label{mutauexp}
\end{align}
Note that another series expansion up to first order with a different
parametrization of the initial neutrino flux ratios for the neutrino
oscillation probabilities has been presented in
Ref.~\cite{Rodejohann:2006qq,Xing:2006uk}. Now, some comments of the
above formulas are in order. First, we discuss the flux ratio
$R_{e\mu}$. In Eq.~(\ref{emuexp}), we have clearly separated first-
and second-order terms in the series expansion. The approximate
relation is obviously much more accurate if referred to small values
of $\theta_{13}$, $\delta_{12}$, and $\delta_{23}$. In order to check
how good our series expansion is, we show in Fig.~\ref{comparison} the
ratio between the approximate formulas (at first and second order,
$R^{\rm first}_{e\mu}$ and $R^{\rm second}_{e\mu}$, respectively) and
the exact formula for $R_{e\mu}$, as a function of the small and
unknown mixing angle $\theta_{13}$. Furthermore, note that the
accidental sum-rule $R_{e\mu} - R_{e\tau} + R_{\mu\tau} = 1$ holds up
to second order in perturbation theory.

In order to fix some values for the fundamental neutrino parameters,
we observe that the following 99~\% C.L. limits hold \cite{Strumia:2006db}
\begin{equation}
\left\{ \begin{array}{rll}
30^\circ < &\theta_{12}& < 38^\circ\\
36^\circ < &\theta_{23}& < 54^\circ\\
0 < &\theta_{13}& < 10^\circ\\
\end{array} \right.
\label{uncertainties}
\end{equation}
with best-fit values corresponding to $\bar \theta_{12}=33^\circ$, 
$\bar \theta_{23}=45^\circ$, and $\bar \theta_{13}=0$.
Thus, in order to be sensitive to $\delta_{12}$ and $\delta_{23}$, we choose 
$\theta_{12}=38^\circ$ and $\theta_{23}=50^\circ$ (corresponding to
$\delta_{12}=\delta_{23}=5^\circ$).

It can be clearly seen that the second-order approximation reproduces
the exact values at the level of 1~\% to 3~\%, the worst case being
obtained for larger values of the mixing angle $\theta_{13}$. The
larger discrepancy between the first-order approximation and the
reference value is to be mainly ascribed to both the relatively large
$\delta_{23}$, which means that second-order terms in this quantity
cannot be safely neglected, and the absence of terms of ${\cal
O}(\delta_{12})$.

At first order in the small parameters, one important feature of this
ratio is that it is independent of the error of $\theta_{12}$, which
only enter in higher-order terms coupled with the error of the mixing
angle $\theta_{23}$. This means that, once you have fixed the value of
the mixing angle $\theta_{12}$, its uncertainties are not relevant for
the determination of $ R_{e\mu}$. Thus, this ratio mainly depends on
$\theta_{13}$ and the uncertainty of $\theta_{23}$.

In order to estimate which of these parameters encodes the largest part of the
uncertainty of $R_{e\mu}$, we can evaluate the maximum spread due to the
combination 
$x=\cos (\delta)\,\theta_{13}$ ($x=\theta_{23}$) at fixed $\theta_{23}$ 
($\cos (\delta)\,\theta_{13}$), namely
\begin{equation}
\Delta R_{e\mu}=\frac{R_{e\mu}(x_{\max})-R_{e\mu}(x_{\min})}{R_{e\mu}(x=0)},
\end{equation}
$x_{\max{},\min{}}$ being the maximum and minimum value of the
variable $x$. It is simple to derive $\Delta R_{e\mu}$ in the
approximations used above. For $x=\cos (\delta)\,\theta_{13}$,
$x_{{\max},{\min}}=\pm \pi/18$ [see Eq.~(\ref{uncertainties})], and
for vanishing $\delta_{12}$ and $\delta_{23}$, we simply obtain
\begin{equation}
\Delta R_{e\mu}=\frac{\pi}{12}\,\sin^2(4\theta_{12})\sim 20~\%,
\end{equation}
for $\theta_{12}$ at the best-fit value. 
On the other hand, for $x=\delta_{23}$, $x_{{\max},{\min}}=\pm
9\,\pi/180$, we obtain
\begin{equation}
\Delta R_{e\mu}=\frac{3\pi}{20}\,\sin^2(2\theta_{12})\sim 45~\%,
\end{equation}
which means that the error of $\theta_{23}$ contains the largest part of the
uncertainty of $R_{e\mu}$ \cite{Fogli:2006jk}.

Then, we discuss the flux ratio $R_{e\tau}$. Due to the approximate
$\mu-\tau$ symmetry
\cite{Fukuyama:1997ky,Mohapatra:1998ka,Ma:2001mr,Lam:2001fb,Harrison:2002et}
(for an incomplete list of references, see Ref.~\cite{Rodejohann:2006qq}),
the flux ratio $R_{e\tau}$ shares similar characteristics as
$R_{e\mu}$, as it can be seen in Eq.~(\ref{etauexp}). In particular,
the first-order approximations are exactly the same, whereas
differences appear in terms of ${\cal O}(\theta^2_{13})$, ${\cal
O}(\delta^2_{23})$, and ${\cal O}(\theta_{13}\,\delta_{23})$.

Finally, we discuss the flux ratio $R_{\mu\tau}$. This flux ratio is
very peculiar. At first order in small parameters, it is exactly 1 and
the deviation from 1 is only present when second-order terms are
included. However, no dependence on the uncertainty $\delta_{12}$ is
present at least up to ${\cal O}(\delta^2_{12})$, which means that
$R_{\mu\tau}$ is mainly sensitive to the central value of the mixing
angle $\theta_{12}$, but, once fixed, the uncertainty of it does not
affect the flux ratio. What is also an important feature,
$R_{\mu\tau}$ is always larger than 1, as it can be seen considering
that the terms in Eq.~(\ref{mutauexp}) under the conditions in
Eq.~(\ref{uncertainties}) cannot give negative corrections to 1. This
feature is not lost in the exact evaluation, as we have numerically
checked. In summary, the flux ratio $R_{\mu\tau}$ is very close to 1,
even if the uncertainties of the neutrino mixing parameters are taken
into account, since the deviation from its standard value is an effect
of second order in small quantities. For that reason, if some ``new
physics'' mechanism would be able to produce a large deviation from 1,
the $R_{\mu\tau}$ flux ratio could give us an interesting possibility
to study such a ``new physics'' scenario.

\begin{figure}
\centerline{\includegraphics[width=0.65\textwidth]{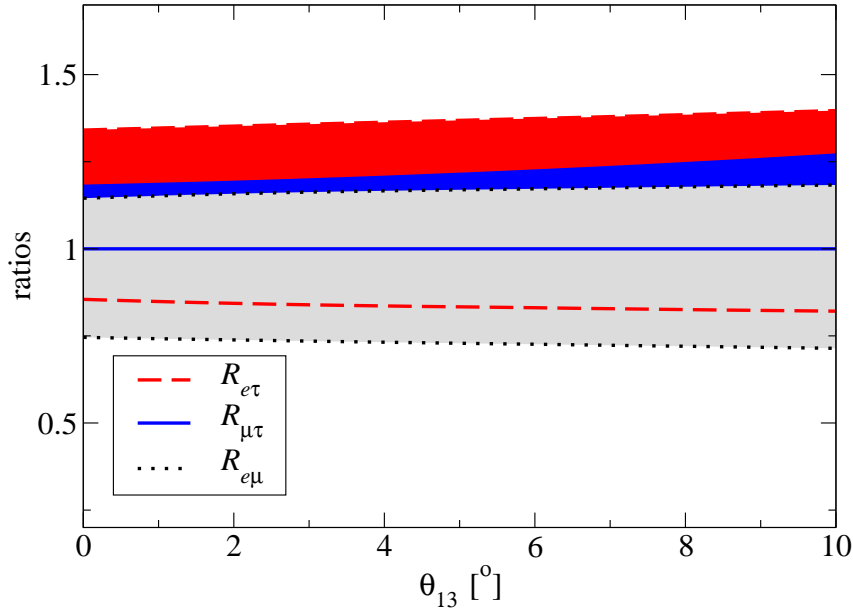}}
\caption{\it The spread of the three flux ratios considered in this
work due to the present uncertainties of the neutrino mixing
parameters [see Eq.~(\ref{uncertainties})], around the best-fit values
$\bar \theta_{12}=33^\circ$ and $\bar \theta_{23}=45^\circ$, as a
function of the mixing angle $\theta_{13}$.  {}From above to below:
$R_{e\tau}$ (red), $R_{\mu\tau}$ (blue), and $R_{e\mu}$ (grey).}
\label{totalunc}
\end{figure}

In order to summarize the previous discussion about uncertainties, we
have evaluated the whole spread of the flux ratios $R_{e\mu}$,
$R_{e\tau}$, and $R_{\mu\tau}$ due to the present lack of knowledge of the
neutrino mixing parameters. We have computed the minimum and maximum
values of each $R_{\alpha\beta}$ varying $\theta_{12}$, $\theta_{23}$,
and $\delta$ in the 99~\% C.L. of Eq.~(\ref{uncertainties}) and
presented the results in Fig.~\ref{totalunc} as a function of
$\theta_{13}$.
{}From above to below, colored areas contain the possible values for
$R_{e\tau}$ (red), $R_{\mu\tau}$ (blue), and $R_{e\mu}$ (grey),
respectively. As it can be easily seen, $R_{e\mu}$ can be as large
(small) as 1.18 (0.72), whereas $R_{e\tau}$ can assume values ranging
from 0.83 to 1.40. On the other hand, $R_{\mu\tau}$ can deviate from
unity by 25~\%.

\begin{figure}
\centerline{\includegraphics[width=0.6\textwidth]{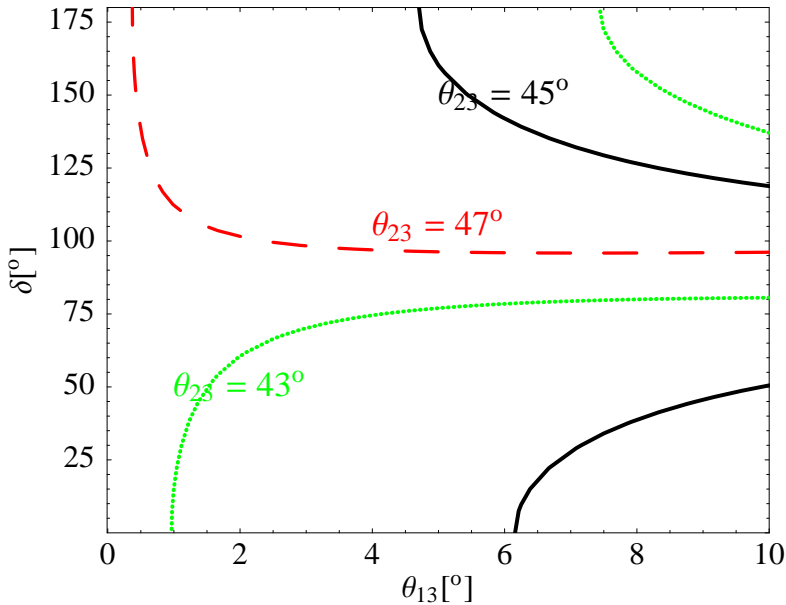}}
\caption{\it The parameter space giving a value $R_{e\mu}=1\pm
5~\%$. The dotted, solid, and dashed curves correspond to
$\theta_{23}=43^\circ,45^\circ,47^\circ$, respectively. The mixing
angle $\theta_{12}$ is fixed to $33^\circ$.}
\label{thetadelta}
\end{figure}

Thus, it seems to be clear that even in the case that the flux ratios
could be measured with infinite precision at neutrino telescopes, it
will be very difficult to estimate or put any reasonable bounds on the
neutrino mixing angles (and the CP-violating phase\footnote{In
Ref.~\cite{Farzan:2002ct}, the possibility of measuring the
CP-violating phase $\delta$ using neutrinos from far distance was
discussed.}) in the standard three-flavor oscillation
framework.\footnote{Note that a first criticism about the possibility
to measure $\theta_{13}$ and $\delta$ at neutrino telescopes can be
found in Ref.~\cite{Vissani:2006pi}.} In particular, the ratios do not
depend in a dramatic way on $\theta_{13}$ (both Figs.~\ref{comparison}
and \ref{totalunc} numerically confirm this statement) and the
dependence on $\delta$ is completely overshaded by the uncertainty of
the mixing angle $\theta_{23}$. However, assuming that some
long-baseline neutrino experiments gave us a very precise measurement
of the mixing angles $\theta_{12}$ and $\theta_{23}$ (which will
certainly not happen in ten years from now) and assuming also the
quite unrealistic scenario in which the flux ratio $R_{e\mu}$ will be
measured within 5~\% precision at future neutrino telescopes,
$\theta_{13}$ and $\delta$ will not be strongly constrained. In
Fig.~\ref{thetadelta}, we quantify this statement showing the allowed
pairs $(\theta_{13},\delta)$, which give $R_{e\mu}=1\pm 5$~\%, for
different values of $\theta_{23}$, namely $43^\circ$, $45^\circ$, and
$47^\circ$, represented by dotted, solid, and dashed curves,
respectively. The admitted parameter space is the region on the
left-hand side of the curves. In principle, some values of
$\theta_{13}$ and $\delta$ can be excluded, depending on the assumed
value of the mixing angle $\theta_{23}$, but the precision obtained
cannot be comparable with that usually claimed for neutrino factories
and $\beta$-beams.
However, as already discussed in Ref.~\cite{Winter:2006ce}, the
combination of the information coming from neutrino telescopes and
those from future neutrino experiment could help in constraining some
of the neutrino mixing parameters, but only in the optimistic hypothesis that
one can clearly separate the different neutrino sources ({\it i.e.}, if the
flux of high-energy neutrinos is under control).

\section{Flux ratios including neutrino decay}
\label{sec:decay}

Neutrino decay has been invoked in the past to solve both the solar
and atmospheric neutrino data problems (see {\it
e.g.}~Refs.~\cite{Pakvasa:1999nh,Lipari:1999vh,Barger:1999bg}). Decay will
deplete the flux of neutrinos on Earth by the exponential factor
\begin{equation}
\label{expfactor}
\exp\left(-\frac{t}{\tau_{\rm lab}}\right) = 
\exp\left(-\frac{L}{E} \frac{m}{\tau}\right)
\end{equation}
in which $\tau$ is the rest-frame neutrino life-time, $L$ the distance
between the source and the detector, and $E$ and $m$ are the neutrino
energy and mass, respectively. Since neutrino decay scenarios have not
been validated so far \cite{Ashie:2004mr,Araki:2004mb}, we can roughly
estimate the order of magnitude of the allowed values of the ratio
$\tau/m$ requiring that $\frac{L}{E} \frac{m}{\tau} \ll 1$. Thus,
giving a lower bound $\tau/m \geq 10^{-4}$~s/eV, obtained for a
typical $L/E$ solar neutrino \cite{Beacom:2002cb}. Note that from the
supernova SN1987A neutrino events, one would obtain a stronger lower
bound at the level of $\tau/m \geq {\cal O}(10^{5})$ s/eV. However,
this bound is not reliable if, as it is the case, the value of
$\theta_{12}$ is large \cite{Frieman:1987as,Fogli:2004gy}.
Given the bounds previously discussed, we cannot eliminate the
possibility of astrophysical neutrino decay. 

\begin{figure}
\centerline{\includegraphics[width=0.6\textwidth]{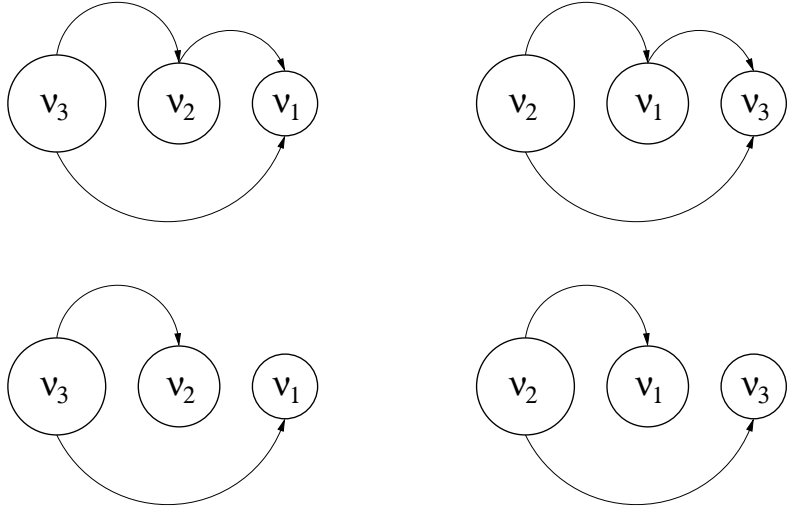}}
\caption{{\it The two different neutrino decay scenarios. Upper-left
  picture: Neutrino decay I with normal hierarchy {\rm ($\alpha_{11} =
  0 \Rightarrow d_1 = 1$)}. Upper-right picture: Neutrino decay I with
  inverted hierarchy {\rm ($\alpha_{33} = 0 \Rightarrow d_3 =
  1$)}. Lower-left picture: Neutrino decay II with normal
  hierarchy {\rm ($\alpha_{11} = \alpha_{22} = 0 \Rightarrow d_1 = d_2
  = 1$)}. Lower-right picture: Neutrino decay II with inverted
  hierarchy {\rm ($\alpha_{11} = \alpha_{33} = 0 \Rightarrow d_1 = d_3
  = 1$)}.}}
\label{fig:neutrinodecay}
\end{figure}

Two-body modes of Majoron type \cite{Chikashige:1980qk,Gelmini:1980re}
are still viable neutrino decay processes, since they are not strongly
constrained (contrary, for example, to radiative decay modes or
three-body modes, see {\it
e.g.}~Refs.~\cite{Acker:1991ej,Pakvasa:1999ta} and references
therein). Stringent bounds on this class of models have been derived
in Ref.~\cite{Hannestad:2005ex}, although the robustness of their
conclusion has been questioned in Ref.~\cite{Bell:2005dr}. Thus, we
retain the possibility for neutrino decay in Majoron models to be a
possible mechanism for flux ratio modifications. We adopt here the
simplifying assumptions that the neutrinos completely decay into the
lightest mass eigenstate or the lightest and next-to-lightest mass
eigenstates over very large $L/E$ and that there are no detectable
decay products. Two different neutrino decay scenarios (with three
active neutrino flavors) will be investigated. In the first scenario,
which will be called ``neutrino decay I'', the two heaviest neutrino
mass eigenstates can decay and only the lightest neutrino mass
eigenstate is stable, whereas in the second scenario, which will be
called ``neutrino decay II'', only the heaviest neutrino mass
eigenstate can decay and the two lightest neutrino mass eigenstates
are stable. The first neutrino decay scenario, neutrino decay I, has
been heavily discussed in the literature, see {\it
e.g.}~Ref.~\cite{Lindner:2001fx}. Note that both neutrino decay
scenarios can be considered for both normal (NH) and inverted neutrino
mass hierarchy (IH), {\it i.e.}, for the cases, $m_1 < m_2 < m_3$,
where $m_1$ is the mass of the lightest neutrino mass eigenstate, and
$m_3 < m_1 < m_2$, where $m_3$ is the lightest neutrino mass
eigenstate. The different neutrino decays scenarios are schematically
shown in Fig.~\ref{fig:neutrinodecay}.

In the case of neutrino decay scenarios, the parameter $\beta = 1$,
and therefore, we can use Eq.~(\ref{eq:<Pab>eff}) in order to compute
averaged neutrino oscillation probabilities (including damping
factors). Note that in the case of neutrino decay, the damping
coefficient matrix is described by neutrino decay parameters
\cite{Blennow:2005yk}, which are defined as $\alpha_{ij} = (\alpha_i +
\alpha_j)/2$, where $\alpha_i = m_i/\tau_i$ is the decay rate with
$m_i$ being the mass of the $i$th mass eigenstate and $\tau_i$ is its
life-time (in its own rest frame) \cite{Lindner:2001fx}. Thus, in the
case of ``neutrino decay I'', we have $\alpha_{11} = 0$ or
$\alpha_{33} = 0$ and all other $\alpha_{ij}$'s are non-zero in
general, since only $\alpha_1 = 0$ (NH) or $\alpha_3 = 0$ (IH), which
means that $d_1 = 1$ or $d_3 = 1$, {\it i.e.}, only one of the $d_i$'s
is equal to 1, and the others are zero (or, in other words, that the
exponential factor in Eq.~(\ref{expfactor}) is unity for the lightest
mass eigenstate and vanishing for the others). In the case of
``neutrino decay II'', we have $\alpha_{11} = \alpha_{22} = 0$ or
$\alpha_{11} = \alpha_{33} = 0$, since $\alpha_1 = \alpha_2 = 0$ (NH)
or $\alpha_1 = \alpha_3 = 0$ (IH), which means that two of the $d_i$'s
are different from zero, {\it i.e.}, $d_1 = d_2 = 1$ or $d_1 = d_3 =
1$. It is now easy to obtain flux ratios on Earth for both scenarios.

\subsection{Neutrino decay I}

The transition probabilities needed to build the flux ratios for the
``neutrino decay I'' scenario, illustrated in the first row of
Fig.~\ref{fig:neutrinodecay}, are obtained from
Eq.~(\ref{eq:<Pab>eff}) in which the terms $d_{i}
J_{\alpha\beta}^{ii}$ have to be evaluated in the unique index
corresponding to the stable mass eigenstate ($i=1$ for NH and $i=3$
for IH). 

For the sake of simplicity and with the aim of checking that our damping
formalism reproduces the results obtained with the standard formulation of
large $L/E$ behavior of the flux ratios, we report here
only the first-order approximations.\footnote{In
Table~\ref{tab:ratios}, the goodness of our approximate formulas can
be found.} Note that the flux ratios are independent of the initial flavor
composition at the source, and according to Ref.~\cite{Beacom:2002vi},
we find that $\phi_{\nu_e}:\phi_{\nu_\mu}:\phi_{\nu_\tau} = |U_{e1}|^2
: |U_{\mu 1}|^2 : |U_{\tau 1}|^2$ for the NH and
$\phi_{\nu_e}:\phi_{\nu_\mu}:\phi_{\nu_\tau} = |U_{e3}|^2 : |U_{\mu
  3}|^2 : |U_{\tau 3}|^2$ for the IH.

First, let us discuss the NH case:
\begin{align}
\label{remund1nh}
R_{e\mu} &= 2 \cot^2\left(\theta_{12}\right)\nonumber\\
&-4 \cot \left(\theta_{12}\right) \csc^2\left(\theta_{12}\right) 
\delta_{12}-4 \cos (\delta ) \cot^3\left(\theta_{12}\right) \theta_{13}+4 
\cot^2\left(\theta_{12}\right) \delta_{23},\\
R_{e\tau} &= 2 \cot^2\left(\theta_{12}\right)\nonumber\\
&-4 \cot \left(\theta_{12}\right) \csc^2\left(\theta_{12}\right) \delta_{12}+4 
\cos (\delta ) \cot^3\left(\theta_{12}\right) \theta_{13}-4 
\cot^2\left(\theta_{12}\right) \delta_{23},\\
R_{\mu\tau} &= 1+4 \cos (\delta ) \cot \left(\theta_{12}\right) \theta_{13}-4 
\delta_{23}.
\end{align}
The main feature of these formulas are, except for the flux ratio
$R_{\mu\tau}$, that the zeroth-order approximation is different from
$1$, being dependent on $\cot^2\left(\theta_{12}\right)$. For the
best-fit value used in this work, this means that
$\phi_{\nu_e}:\phi_{\nu_\mu}:\phi_{\nu_\tau}\sim 5:1:1$
\cite{Beacom:2002vi}, a very large deviation compared to the standard
neutrino framework. Moreover, unlike Eqs.~(\ref{emuexp}) and
(\ref{etauexp}), a dependence on $\delta_{12}$ appears at first order,
which means that the error of the mixing angle $\theta_{12}$ cannot be
neglected. Note that $R_{\mu\tau}$ obtains corrections with respect to
1 also at first order. Thus, we expect that the intrinsic uncertainty
of the flux ratios (given by our ignorance on the neutrino mixing
parameters) is much larger than the standard framework case.

Second, in the case of IH, the flux ratios assume the following simple
structure
\begin{align}
\label{remund1ih}
R_{e\mu}&= R_{e\tau}= 0,\\
R_{\mu\tau} &= 1 + 4 \delta_{23}.
\end{align}
The signature of this particular case is extremely different from both
the standard framework and the ``neutrino decay I'' with
NH. Corrections to $R_{e\mu}= 0$ and $R_{e\tau}= 0$ appear at second
order in small quantities 
and are
dependent on $\theta_{13}$. Thus, we expect a substantial deviation
from zero at large $\theta_{13}$, since corrections appear only at
second order in $\theta_{13}$ \cite{Beacom:2003zg}.

\subsection{Neutrino decay II}

Similarly, the ``neutrino decay II'' scenario is depicted in the
second row in Fig.~\ref{fig:neutrinodecay}. In this case, the two
lightest neutrino mass eigenstates are stable instead of only the
lightest in the ``neutrino decay I'' scenario. Therefore, we will have
two $d_i J_{\alpha\beta}^{ii}$ terms in Eq.~(\ref{eq:<Pab>eff})
instead of only one ($i = 1$ and $i=2$ for NH and $i = 1$ and $i=3$
for IH). In App.~\ref{sub:1st-order}, we present formulas for the flux
ratios only up to first order in the small parameters $\theta_{13}$,
$\delta_{12}$, and $\delta_{23}$, since the first-order formulas make
a much better approximation for ``neutrino decay II'' than ``neutrino
decay I''. Note that for NH, the formulas are independent of
$\delta_{12}$. In addition, for both NH and IH, all formulas are
dependent on $\delta_{23}$, which means that the uncertainty of
$\theta_{23}$ is crucial. Furthermore, we observe that the structure
of the formulas for NH and IH is very different. For IH, nearly all
terms (except from the zeroth-order term of the flux ratio
$R_{\mu\tau}$) include the factor $3 - \cos(2\theta_{12}) = 2 +
2 \sin^2(\theta_{12})$ in the denominators, which has a value between 2
and 4 and is therefore not equal to zero at any time though. In
addition, $R_{e\mu}$ is independent of the CP-violating phase
$\delta$, whereas $R_{e\tau}$ is dependent on $\delta$, which actually
spoils the approximate $\mu-\tau$ symmetry. In some sense, the values
of the flux ratios are closer to the standard neutrino framework for
``neutrino decay II'' than ``neutrino decay I''.\footnote{Again, in
Table~\ref{tab:ratios}, we present the numerical results on the flux
ratios.}

\subsection{Discussion about the neutrino decay scenarios}

\begin{figure}
\centerline{\includegraphics[width=0.51\textwidth,angle=-90,clip]{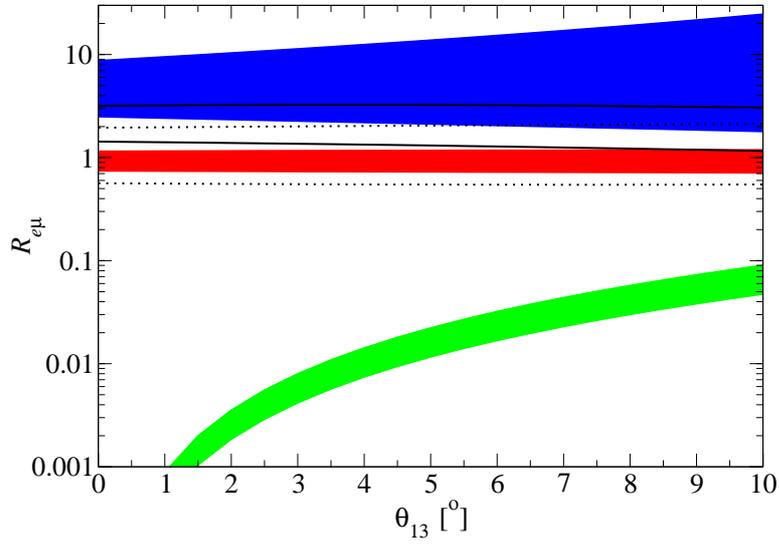}}
\caption{{\it The flux ratio $R_{e\mu}$ including neutrino decay as a
  function of the mixing angle $\theta_{13}$. The red (middle) area
  corresponds to the standard neutrino oscillation framework, whereas
  the blue (upper) and green (lower) areas correspond to neutrino
  decay with the lightest neutrino mass eigenstate being stable only
  for NH and IH, respectively. The case of the two lightest neutrino
  mass eigenstates being stable is visualized by the solid (NH) and
  dotted (IH) curves, respectively. The uncertainties of the neutrino
  mixing angles are those given in Eq.~(\ref{uncertainties}), around
  the best-fit values $\bar \theta_{12}=33^\circ$ and $\bar
  \theta_{23}=45^\circ$.}}
\label{fig:Rem}
\end{figure}

\begin{figure}
\centerline{\includegraphics[width=0.51\textwidth,angle=-90,clip]{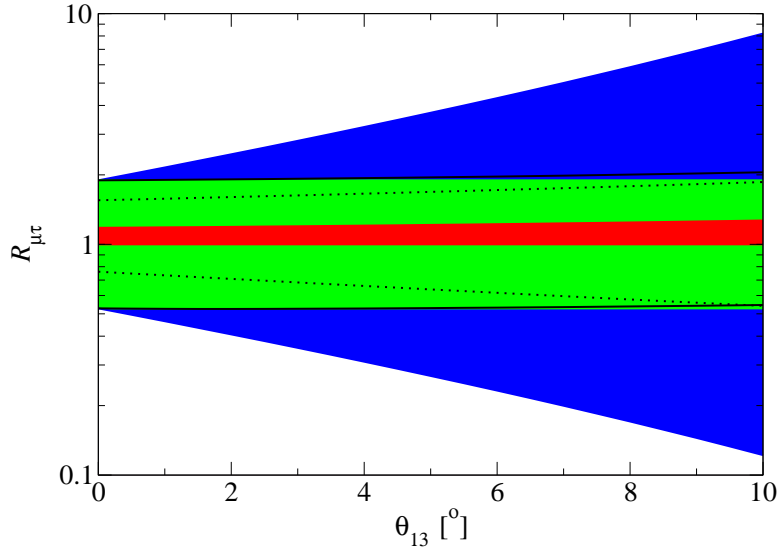}}
\caption{{\it The flux ratio $R_{\mu\tau}$ including neutrino decay as
  a function of the mixing angle $\theta_{13}$. The red (innermost)
  area corresponds to the standard neutrino oscillation framework,
  whereas the blue (outermost) and green (middle) areas correspond to
  neutrino decay with the lightest neutrino mass eigenstate being
  stable only for NH and IH, respectively. The case of the two
  lightest neutrino mass eigenstates being stable is visualized by the
  solid (NH) and dotted (IH) curves, respectively. The uncertainties
  of the neutrino mixing angles are those given in
  Eq.~(\ref{uncertainties}), around the best-fit values $\bar
  \theta_{12}=33^\circ$ and $\bar \theta_{23}=45^\circ$.}}
\label{fig:Rmt}
\end{figure}

The previous considerations have been summarized in
Figs.~\ref{fig:Rem} and \ref{fig:Rmt}, in which we present the flux
ratios $R_{e\mu}$ and $R_{\mu\tau}$, respectively, using the two
neutrino decay scenarios with both NH and IH\footnote{Note that the
flux ratio $R_{e\tau}$ is very similar to the flux ratio $R_{e\mu}$,
and we have therefore not presented this flux ratio in a separate
figure.}.

In Fig.~\ref{fig:Rem} for the flux ratio $R_{e\mu}$, we observe that
the ``neutrino decay I'' scenario for both NH and IH can, in
principle, be distinguished from the standard neutrino oscillation
framework. As expected from the very different first-order
approximations in Eqs.~(\ref{remund1nh}) and (\ref{remund1ih}), even
including the errors of the neutrino mixing parameters, the bands
representing the total spread of $R_{e\mu}$ are very well separated,
especially for the IH case in which 
a quadratic behavior of
$\theta_{13}$ 
can be clearly seen
(notice that we are using a log scale on the vertical axis). In
addition, we observe the large dependence of $\theta_{13}$ on the
uncertainty of the NH case, which is mainly to be ascribed to the term
$4 \cos (\delta ) \cot^3\left(\theta_{12}\right) \theta_{13}$.

A very different situation arises for the ``neutrino decay II''
scenario (solid and dotted curves for NH and IH in Fig.~\ref{fig:Rem},
respectively). Due to the fact that this scenario contains two stable
mass eigenstates, it is quite similar to the standard neutrino
oscillation framework, in which all the mass eigenstates are not
allowed to decay. Therefore, differences in $R_{e\mu}$ are very
difficult to be discerned (the same comment can be applied to
Fig.~\ref{fig:Rmt}).

In Fig.~\ref{fig:Rmt}, the results of the flux ratio $R_{\mu\tau}$ are
shown. As already outlined in the previous section, in the standard
framework, $R_{\mu\tau}$ is always larger than 1, even including the
uncertainties of the neutrino mixing angles. The neutrino decay
scenarios allow $R_{\mu\tau}$ to be smaller than 1 (which could be a
clear signature of ``new physics'' effects), but the large spread of
the ratio at any value of $\theta_{13}$ does not allow an explicit
separation between the standard and decay phenomenologies. Thus, this
flux ratio is not a good candidate to find deviations from the
standard neutrino oscillation framework.

In Table~\ref{tab:ratios}, we present numerical calculations of the
flux ratios $\phi_{\nu_e}:\phi_{\nu_\mu}:\phi_{\nu_\tau}$ for the
different neutrino decay scenarios, comparing the zeroth-, first-, and
second-order results with exact calculations.\footnote{Note that the
zeroth-order values for the flux ratios in the case of ``neutrino
decay I'' with both NH and IH can, {\it e.g.}, also be found in
Ref.~\cite{Beacom:2002vi}.} In general, we observe that the
second-order results contains all the features connected with
non-maximal $\theta_{23}$ and non-vanishing $\theta_{13}$. Thus, it
seems to be important to include second-order effects in any
analytical treatment of flux ratios. However, note that the
first-order results work especially well for the standard framework
and ``neutrino decay II'', whereas for ``neutrino decay I'', the
first-order results do not reproduce all features of the exact
computations.
\begin{table}
\begin{center}
\begin{tabular}{l|c|c|c|c}
\hline\hline
 &  & & & \\[-2.5mm]
\hspace{1cm} $\phi_{\nu_e}:\phi_{\nu_\mu}:\phi_{\nu_\tau}$
 & $0^{\rm th}$ & $1^{\rm st}$ & $2^{\rm nd}$ & exact \\ 
 &  & & & \\[-2.5mm]
\hline
 &  & & \\[-2.5mm]
standard framework& 1:1:1 & 0.93:1:1&0.93:1.06:1 & 0.93:1.06:1\\ 
 &  & & &\\[-2.5mm]
\hline
 &  & & &\\[-2.5mm]
neutrino decay I (NH)& 4.74:1:1 & 3.08:1.06:1&3.14:0.99:1 &3.20:0.99:1\\ 
 &  & & &\\[-2.5mm]
\hline 
 &  & & &\\[-2.5mm]
neutrino decay I (IH)& 0:1:1 & 0:1.35:1&0.02:1.41:1\ &0.02:1.42:1\\
 &  & & &\\[-2.5mm]
\hline 
 &  & & &\\[-2.5mm]
neutrino decay II (NH)& 2:1:1 & 1.81:0.65:1& 1.78:0.74:1&1.78:0.73:1\\
 &  & & &\\[-2.5mm]
\hline 
 &  & & &\\[-2.5mm]
neutrino decay II (IH)&
1.08:1:1 & 0.91:1.28:1& 0.93:1.30:1&0.93:1.30:1\\[-2.5mm]
 &  & & &\\
\hline\hline 
\end{tabular}
\end{center}
\caption{
\label{tab1} \it Flux ratios
$\phi_{\nu_e}:\phi_{\nu_\mu}:\phi_{\nu_\tau}$ for the different
neutrino decay scenarios. $0^{\rm th}$ , $1^{\rm st}$, and $2^{\rm
nd}$ represent the results obtained up to zeroth, first, and second
order in small parameters, respectively, whereas the exact computation
is given in the column {\rm exact}. The values of the fundamental
neutrino parameters used are: $\theta_{12} = 33^\circ$, $\theta_{13} =
5^\circ$, and $\delta = 40^\circ$. In addition, we have assumed
$\delta_{12} = 5^\circ$ and $\delta_{23} = 5^\circ$.}
\label{tab:ratios}
\end{table}

\subsection{Discussion about neutrino telescopes}
\label{neutel}

In the previous sections, we have discussed the flux ratios from a
theoretical point of view; thus taking into account the role of the
uncertainties of the neutrino mixing parameters. Let us now address
the question of what can really be measured at neutrino telescopes and
how the {\it statistics}, {\it i.e.}, the number of expected events,
affects the results that can be obtained.

The measurement of neutrino flavor seems to be a very difficult task,
which means that the three flux ratios previously studied are unlikely
to be directly accessible. However, they can be inferred from distinct
classes of events: showers, muon tracks, and unique $\nu_\tau$
signatures (see, {\it e.g.},~Ref.~\cite{beacom2003}). All neutrino
flavors undergo neutral-current interactions (inside or nearby the
detector), which result in hadronic showers in a detector. On the
other hand, $\nu_e$ charged-current interactions also produce
electromagnetic showers that, in principle, could be distinguished
from the hadronic ones due to the different muon content (which is
absent in the electromagnetic showers). The shower rate also includes
$\nu_\tau$ charged-current events, since at least for energies below
the order of PeV the tau track cannot be separated from the
shower. For energies above the order of PeV, double-bang
\cite{Learned:1994wg,Athar:2000rx} and lollipop events are also
possible, but we do not discuss these tau signatures in this
work. Muon tracks originate from $\nu_\mu$ charged-current
interactions; muons always emerge from the shower, which means that
this kind of process can be distinguished from a typical shower event.

Neutrino telescopes seem to be the ideal place where to seek for
high-energy neutrino interactions. In deep ice or water, neutrinos are
detected by observation of the Cherenkov light emitted by charged
particles produced in charged-current and neutral-current
interactions. Such detectors are mainly sensitive to TeV-PeV neutrino
energies; thus opening the possibility of studying this still
unexplored energy regime.  In order to observe a hadronic event inside
a telescope, a neutrino must first {\it survive} as it crosses the
atmosphere and then the ice (or water) above the detector. Its typical
interaction length in a medium of density $\rho$ is $ L_0=1/(\rho
N_A\sigma^{\nu N})$, where $\sigma^{\nu N}=\sigma^{\nu N}_{\rm
CC}+\sigma^{\nu N}_{\rm NC}$ is the total cross-section to have an
interaction that depletes the neutrino energy. It is usual to express
this length in terms of its depth: $x_0=\rho L_0$ ({\it i.e.}, one
meter of water has a depth of 100 g/cm$^2$).

A neutrino from a zenith angle $\theta_z$ must cross a column density
of material $x(\theta_z)=\int_{\theta_z} {\rm d}l\ \rho(l,\theta_z)$.
In practice, the path in the atmosphere is negligible and
$x(\theta_z)$ is just the depth of the ice or water above the
detector. The probability that it does not interact before reaching
the detector is then $P_{\rm surv}(E_\nu,\theta_z)={\rm
e}^{-x/x_0}$. Once in the detector, the probability of an event is
$P_{\rm int}(E_\nu)\approx 1-{\rm e}^{-L\rho N_A\sigma^{\nu N}_{\rm
int}}$, where $L$ is the linear dimension of the detector and
$\sigma^{\nu N}_{\rm int}$ is the cross-section for the interaction
taken into account (charged or neutral). Therefore, the number of
shower events in the telescope in an observation time $T$ for a fixed
flavor is
\begin{eqnarray}
\label{eventi}
N_{\rm sh}=2\pi AT \int {\rm d}E_\nu 
\frac{{\rm d}\Phi_{\nu_i}}{{\rm d}E_{\nu}}
\int {\rm d}\cos\theta_z P_{\rm surv} \int_{y_{\min{}}}^{y_{\max{}}}
     {\rm d}y \frac{1}{\sigma_{\nu N}}\, 
\frac{{\rm d}\sigma_{\nu N}}{{\rm d}y}\,P_{\rm int},
\end{eqnarray}
where $A$ is the detector's cross-sectional area, ${\rm
d}\Phi_{\nu_i}/{\rm d}E_\nu$ is the neutrino flux, and $\theta_z$ is
the zenith angle. The variable $y$ is the usual inelasticity parameter
in deep-inelastic scattering and represents the fraction of the
neutrino energy going into hadronic channels, which means that $E_{\rm
sh}=y\, E_\nu$.

According to our previous discussion about flavor detectability, for
neutral-current interactions and $\nu_\tau$ charged-currents (for
energies below a few PeV), we should consider $y_{\min{}}=y_{\rm
thr}=E_{\rm sh}^{\rm thr}/E_\nu$ and $y_{\max{}}=1$, where $E_{\rm
sh}^{\rm thr}$ is the threshold energy for shower detection. For
$\nu_e$ charged-currents, $y_{\min{}}=0$ and $y_{\max{}}=1$, since the
outgoing electron also shower. Moreover, for $\nu_\tau$
Eq.~(\ref{eventi}) should be modified to include the effects of
regeneration. In fact, because of the short lifetime of tau, the
$\nu_\tau\to \tau\to \nu_\tau$ conversion takes place, with the result
of softening the neutrino energy. In principle, one should perform a
dedicated analysis to determine the average energy a $\nu_\tau$ has
when it reaches the detector. Such a study is beyond the scope of our
work (but see, {\it e.g.}, Refs.~\cite{Halzen:1998be,Dutta:2000jv});
we approximate this fact assuming that, in neutral-current events, the
final neutrino energy is about one half of the initial energy and that,
in charged-current events, the $\nu_\tau$ energy is reduced to about one fifth
\cite{Halzen:1998be}. Thus, we can evaluate these integrals to obtain
the total number of shower events:
\begin{equation}
N^{\rm tot}_{\rm sh}=\sum_{\nu_e,\nu_\mu,\nu_\tau} N^{\rm NC}_{\rm sh}+
\sum_{\nu_e,\nu_\tau} N^{\rm CC}_{\rm sh}.
\end{equation}
Equation~(\ref{eventi}) is a simple form to evaluate the shower rates
at neutrino telescopes. Even if an exhaustive discussion of the
problems related to the evaluation of rates is beyond the scope of
this work, we should mention two main sources of
uncertainties. Obviously, we do not know the neutrino fluxes reaching
the Earth's surface, neither the normalization nor the energy
behavior. Thus, we can assume the form
\begin{equation}
\label{fluxes}
\frac{{\rm d}\Phi_{\nu_i}}{{\rm d}E_{\nu}}= c\, \phi_{\nu_i}\,E_\nu^{-2},
\end{equation}
where $c$ is a normalization constant and $\phi_{\nu_i}$'s are fluxes
entering in the definition of the flux ratios in
Eq.~(\ref{ratios}). When calculating the ratios of rates, the
normalization factor cancels out, but the spectral index is still a
crucial ingredient. In addition, the neutrino-nucleon cross-section
is also affected by large uncertainties, mainly due to the
extrapolation to very small $x$ required to describe the extremely
high-energy regime involved in the processes under discussion.

A high-energy muon undergoes energy loss propagating in the medium,
and thus, generating showers along its track by bremsstrahlung, pair
production, and photonuclear interactions. These showers are sources
of Cherenkov light and can be easily distinguished from showers
previously discussed. Equation~(\ref{eventi}) applies, but now
\begin{eqnarray}
P_{\rm int}(E_\nu)\approx 1-{\rm e}^{-R_\mu\rho N_A\sigma^{\nu N}_{\rm int}},
\label{pint2}
\end{eqnarray}
where
\begin{eqnarray}
R_\mu=\frac{1}{\beta}\,{\rm ln}\left(\frac{\alpha+\beta\,E_\mu}
{\alpha+\beta\,E^{\rm thr}_\mu}\right)
\end{eqnarray}
with $\alpha=2$~MeV~cm$^2$/g, $\beta=4.2\times 10^{-6}$~cm$^2$/g,
$E_\mu=(1-y)\,E_\nu$, and $E^{\rm thr}_\mu$ is the value of the muon energy
to be detected. Note that the $y$ integration in Eq.~(\ref{eventi})
is modified in such a way that $y_{\min{}}=0$ and
$y_{\max{}}=y_{\rm thr}^\mu=1-E_\mu^{\rm thr}/E_\nu$.

{}From the previous discussion, it seems to be clear that a viable
experimental quantity measurable at neutrino telescopes is the muon
tracks $N_{\rm tr}$ to shower ratio. Factorizing out $\phi_{\nu_i}$
from Eq.~(\ref{fluxes}) and calculating the expected rates according
to Eq.~(\ref{eventi}), we obtain
\begin{equation}
\label{exprate}
{\cal R}=\frac{N_{\rm tr}}{N_{\rm sh}}=\frac{\phi_{\nu_\mu}\,N_{\rm tr}}
{\sum_{i=e,\tau}\phi_{\nu_i}\,(N_{{\rm sh},\nu_i}^{\rm
    CC}+N_{{\rm sh},\nu_i}^{\rm NC})
+\phi_{\nu_\mu}\,N_{{\rm sh},\nu_\mu}^{\rm NC}}.
\end{equation}
Note that the ratio ${\cal R}$ depends on the neutrino mixing
parameters through the analogous dependence in $\phi_{\nu_i}$. 

\begin{figure}
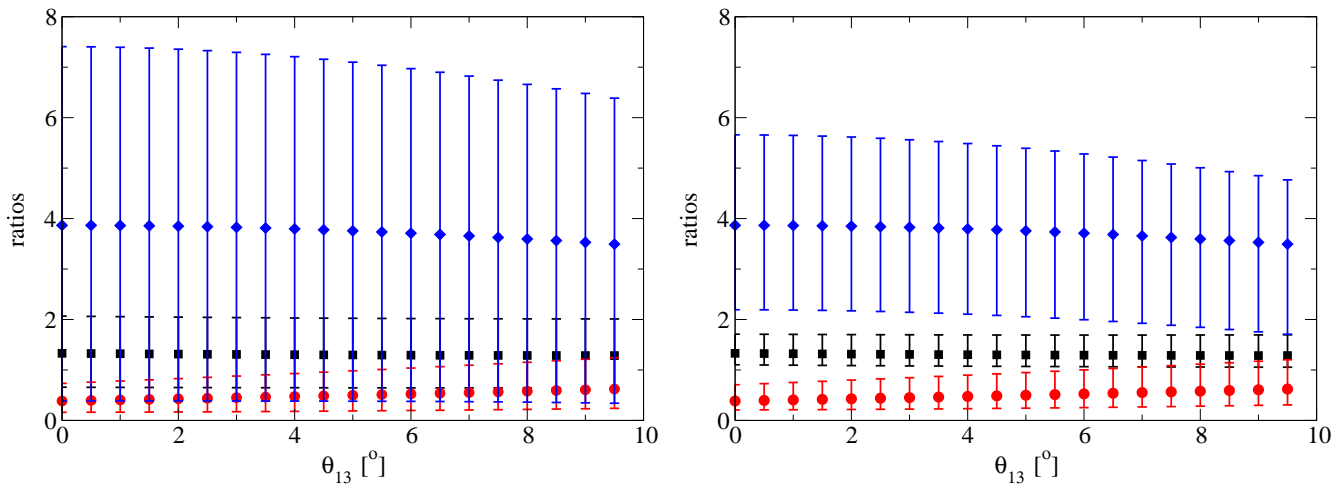

\centerline{\includegraphics[width=0.5\textwidth,clip]{res.eps}
~\includegraphics[width=0.5\textwidth,clip]{res10stat.eps}}
\caption{\it The muon-to-shower ratio ${\cal R}$ as a function of the
mixing angle $\theta_{13}$, evaluated for three of the scenarios
discussed in the main text: standard neutrino framework (black boxes),
``neutrino decay I'', NH (blue diamonds), and ``neutrino decay I'', IH
(red circles). Left plot: Results obtained considering one year of data
taking at IceCube.  Right plot: Results obtained considering ten years
of data taking at IceCube.  In both plots, the uncertainties of the
neutrino mixing angles are those given in Eq.~(\ref{uncertainties}).}
\label{expratio}
\end{figure}

In order to verify these statements, we can use Eq.~(\ref{exprate})
once the energy thresholds for muons track and shower detection have
been specified. To give an example, we use the IceCube detector setup
and assume a quite conservative energy threshold $E_{\rm sh}^{\rm
thr}=E_{\rm \mu}^{\rm thr}=500$~TeV for both types of processes in
order to consider event rates well above the atmospheric neutrino
backgrounds \cite{alvarez}, although for up-going neutrinos the
threshold could be sizably smaller due to the screening effect of the
Earth. For this setup, the result of ${\cal R}$ as a function of
$\theta_{13}$ is shown in Fig.~\ref{expratio}, in which we have
evaluated ratios ${\cal R}$ for the standard neutrino oscillation
scenario and ``neutrino decay I'' (both hierarchies). The error bars
represent the statistical error associated with the number of muon
tracks and shower events [thus, including also the uncertainties of the
mixing parameters given in Eq.~(\ref{uncertainties})].

For neutrino fluxes in the ratio
$\phi_{\nu_e}:\phi_{\nu_\mu}:\phi_{\nu_\tau}=1:1:1$ (standard
scenario), we expect about 15 muon tracks against about 11 shower
events in one year of data taking.  As shown in the left plot of
Fig.~\ref{expratio}, this event rate is not enough to distinguish
among the three scenarios considered in the text. A large overlap
between the standard scenario and ``neutrino decay I'' is present for almost
any value of $\theta_{13}$. We have verified that a large contribution
to the ratio errors is due to poor statistics. If we increase the
number of expected events by a factor of $10$ (that could be achieved
in ten years of data taking), we observe a strong reduction of the
uncertainty of the ratios ${\cal R}$, as illustrated in the right plot
of Fig.~\ref{expratio}. Especially, at smaller $\theta_{13}$, the
ratios corresponding to the three scenarios are quite separated; thus
the sensitivity of IceCube seems to be enough to tell standard from
new physics scenarios involving neutrino decay apart. Note that the effect
we have observed increasing the event rates cannot be mimicked by a
reduction of the uncertainties of the neutrino mixing parameters.  We
have calculated the ratios ${\cal R}$ assuming a 10~\% error for both
$\theta_{12}$ and $\theta_{23}$; we have clearly observed a reduction
of the errors of ${\cal R}$, but not sufficient to avoid the overlap
among different scenarios, which means that, even for very precise
measurements of the solar and atmospheric mixing angles, the error is
dominated by statistics. In the opposite limit, that of infinite
statistics, the right plot suggests (which we have numerically
verified) that the distinction between the scenarios under study is
feasible even with the current uncertainties of $\theta_{12}$ and
$\theta_{23}$. Note that this statement has nothing to do with the
fact that we can learn something on the neutrino mixing angles at
neutrino telescopes, as we have extensively discussed in the previous
sections.

\section{Summary and conclusions}
\label{sec:summary}

The next generation of neutrino telescopes will be able to observe
many high-energy events, thus opening the very exciting possibility to
measure neutrino flux ratios. However, many uncertainties from both
theoretical and experimental point of view, can spoil the capability
of extracting useful information about astrophysical sources and
fundamental neutrino properties. In this work, we have studied in
detail the analytical behavior of the three flux ratios $R_{e\mu}$,
$R_{e\tau}$, and $R_{\mu\tau}$, focusing on how the current errors of
the neutrino mixing parameters can affect the theoretical expectations
of $R_{\alpha\beta}$ for any $\alpha$ and $\beta$ (for flux ratios at
the source
$\phi^0_{\nu_e}:\phi^0_{\nu_\mu}:\phi^0_{\nu_\tau}=1:2:0$). Due to the
enormous distances traveled by neutrinos, the transition probabilities
$P_{\alpha\beta}$ among flavors generally assume a simple analytical
structure. We have derived them starting from a general treatment of
$P_{\alpha\beta}$ which includes damping parameters to account for
possible ``new physics'' effects in neutrino oscillations. We have
showed that the only non-vanishing large $L$ averaged transitions are
those in which the non-diagonal damping factors completely disappear
from the theory. We have then expanded the flux ratios in terms of
small parameters, namely the deviation from maximal $\theta_{23}$, the
deviation of $\theta_{13}$ from zero, and the deviation of
$\theta_{12}$ from its best-fit value. Furthermore, we have studied in
detail the uncertainties of $R_{\alpha\beta}$ connected with our
ignorance of the fundamental neutrino physics. We have observed that
the largest indetermination on $R_{e\mu}$ (at the level of 45~\%) is
due to the fact that we do not know the $\theta_{23}$ octant, whereas
the uncertainty associated with the product $\cos(\delta)
\,\theta_{13}$ is at least a factor of two smaller. Due to the
approximate $\mu-\tau$ symmetry, the same conclusion can be drawn for
$R_{e\tau}$. On the other hand, $R_{\mu\tau}$ behaves in a very
peculiar way, since the corrections from the standard value
$R_{\mu\tau}=1$ are positive and of ${\cal O}$($\delta_{12}^2$). These
considerations have been summarized in Fig.~\ref{comparison}, in which
the total spread of the flux ratios caused by the current
uncertainties of mixing parameters seems to be too large to admit the
possibility of any realistic measurement of fundamental parameters in
the neutrino sector, ({\it e.g.}, $\theta_{13}$, $\delta$, and the
octant of $\theta_{23}$).

We have performed the same detailed study on two different scenarios
involving neutrino decay, considering the lightest mass eigenstate
(neutrino decay I) or the lightest and next-to-lightest (neutrino decay
II) as a stable particle(s). Once the errors of the mixing parameters
are included, we have observed that the most promising ratio to
measure deviations from the standard framework is $R_{e\mu}$,
especially for the ``neutrino decay I'' scenario with inverted
hierarchy. In that case, especially for smaller values of
$\theta_{13}$, the difference between the ratios can be as large as
two orders of magnitude. On the other hand, the ``neutrino decay II''
scenario can be hardly distinguished from the standard result.

Finally, we have studied the muon tracks to shower ratio ${\cal R}$ at
IceCube, as a physical observable directly connected to the
experiments. We have evaluated ${\cal R}$ in the standard picture of
neutrino oscillations and the ``neutrino decay I'' scenario for both
hierarchies. We have found that, contrary to the case in which ${\cal
R}$ is computed with the neutrino mixing parameters fixed to their
best-fit values, the inclusion of parameter uncertainties reduces the
possibility to distinguish ``new physics'' effects in neutrino
oscillations. Thus, we believe that any realistic analysis of physics
reach of neutrino telescopes concerning flux ratios should carefully
include parameters uncertainties.

\subsection*{Acknowledgments}

D.M. wishes to thank Paolo Lipari and Olga Mena for very interesting
discussions. T.O. would like to thank Evgeny Akhmedov, Mattias
Blennow, and Per Olof Hulth for useful discussions. In addition,
D.M. would like to thank for the hospitality of the Royal Institute of
Technology (KTH) where part of this work was done.

This work was supported by the Royal Swedish Academy of Sciences
(KVA), the Swedish Research Council (Vetenskapsr{\aa}det), Contract
No.~621-2005-3588, and the Magnus Bergvall Foundation.

\appendix

\section{Series expansion formulas}
\subsection{First-order series expansions of neutrino flux ratios with two non-zero normalized damping factors}
\label{sub:1st-order}

The neutrino flux ratios using two of the normalized damping parameters
equal to 1, {\it i.e.}, $d_1=d_2=1$, and the last one equal to zero:
\begin{align}
R_{e\mu}(d_1 = d_2 = 1) &= 2 + 2 \cos(\delta) \sin(4\theta_{12}) \theta_{13} + 2 [3 + \cos(4\theta_{12})] \delta_{23}, \\
R_{e\tau}(d_1 = d_2 = 1) &= 2 + 2 \cos(\delta) \sin(4\theta_{12}) \theta_{13} - 4 \sin^2(2\theta_{12}) \delta_{23}, \\
R_{\mu\tau}(d_1 = d_2 = 1) &= 1 - 4 \delta_{23}.
\end{align}
The neutrino flux ratios using two of the normalized damping parameters
equal to 1, {\it i.e.}, $d_1=d_3=1$, and the last one equal to zero:
\begin{align}
R_{e\mu}(d_1 = d_3 = 1) &= \frac{4 \cos^2(\theta_{12})}{3-\cos(2\theta_{12})} - \frac{16 \sin(2\theta_{12})}{[3-\cos(2\theta_{12})]^2} \delta_{12} - \frac{32 \cos^2(\theta_{12})}{[3-\cos(2\theta_{12})]^2} \delta_{23},\\
R_{e\tau}(d_1 = d_3 = 1) &= \frac{4 \cos^2(\theta_{12})}{3-\cos(2\theta_{12})} - \frac{16 \sin(2\theta_{12})}{[3-\cos(2\theta_{12})]^2} \delta_{12} \nonumber\\
&+ \frac{32 \cos(\delta) \cos^3(\theta_{12})\sin(\theta_{12})}{[3-\cos(2\theta_{12})]^2} \theta_{13} - \frac{8 \sin^2(2\theta_{12})}{[3-\cos(2\theta_{12})]^2} \delta_{23},\\
R_{\mu\tau}(d_1 = d_3 = 1) &= 1 + \frac{4 \cos(\delta) \sin(2\theta_{12})}{3-\cos(2\theta_{12})} \theta_{13} + \frac{8 \cos^2(\theta_{12})}{3-\cos(2\theta_{12})} \delta_{23}.
\end{align}


\begin{thebibliography}{10}
\expandafter\ifx\csname bibnamefont\endcsname\relax
  \def\bibnamefont#1{#1}\fi
\expandafter\ifx\csname bibfnamefont\endcsname\relax
  \def\bibfnamefont#1{#1}\fi
\expandafter\ifx\csname url\endcsname\relax
  \def\url#1{\texttt{#1}}\fi
\expandafter\ifx\csname urlprefix\endcsname\relax\def\urlprefix{URL }\fi
\providecommand{\bibinfo}[2]{#2}
\providecommand{\eprint}[2][]{\url{#2}}

\bibitem{Achterberg:2006md}
\bibinfo{author}{\bibfnamefont{A.}~\bibnamefont{Achterberg}} \emph{et~al.}
  (\bibinfo{collaboration}{IceCube}), \bibinfo{journal}{Astropart. Phys.}
  \textbf{\bibinfo{volume}{26}}, \bibinfo{pages}{155} (\bibinfo{year}{2006}),
  \eprint{astro-ph/0604450}.

\bibitem{Katz:2006wv}
\bibinfo{author}{\bibfnamefont{U.~F.} \bibnamefont{Katz}},
  \bibinfo{journal}{Nucl. Instrum. Meth.} \textbf{\bibinfo{volume}{A567}},
  \bibinfo{pages}{457} (\bibinfo{year}{2006}), \eprint{astro-ph/0606068}.

\bibitem{Athar:2000yw}
\bibinfo{author}{\bibfnamefont{H.}~\bibnamefont{Athar}},
  \bibinfo{author}{\bibfnamefont{M.}~\bibnamefont{Je{\.z}abek}},
  \bibnamefont{and} \bibinfo{author}{\bibfnamefont{O.}~\bibnamefont{Yasuda}},
  \bibinfo{journal}{Phys. Rev.} \textbf{\bibinfo{volume}{D62}},
  \bibinfo{pages}{103007} (\bibinfo{year}{2000}), \eprint{hep-ph/0005104}.

\bibitem{Farzan:2002ct}
\bibinfo{author}{\bibfnamefont{Y.}~\bibnamefont{Farzan}} \bibnamefont{and}
  \bibinfo{author}{\bibfnamefont{A.~Y.} \bibnamefont{Smirnov}},
  \bibinfo{journal}{Phys. Rev.} \textbf{\bibinfo{volume}{D65}},
  \bibinfo{pages}{113001} (\bibinfo{year}{2002}), \eprint{hep-ph/0201105}.

\bibitem{Bhattacharjee:2005nh}
\bibinfo{author}{\bibfnamefont{P.}~\bibnamefont{Bhattacharjee}}
  \bibnamefont{and} \bibinfo{author}{\bibfnamefont{N.}~\bibnamefont{Gupta}}
  (\bibinfo{year}{2005}), \eprint{hep-ph/0501191}.

\bibitem{Serpico:2005sz}
\bibinfo{author}{\bibfnamefont{P.~D.} \bibnamefont{Serpico}} \bibnamefont{and}
  \bibinfo{author}{\bibfnamefont{M.}~\bibnamefont{Kachelrie{\ss}}},
  \bibinfo{journal}{Phys. Rev. Lett.} \textbf{\bibinfo{volume}{94}},
  \bibinfo{pages}{211102} (\bibinfo{year}{2005}), \eprint{hep-ph/0502088}.

\bibitem{Serpico:2005bs}
\bibinfo{author}{\bibfnamefont{P.~D.} \bibnamefont{Serpico}},
  \bibinfo{journal}{Phys. Rev.} \textbf{\bibinfo{volume}{D73}},
  \bibinfo{pages}{047301} (\bibinfo{year}{2006}), \eprint{hep-ph/0511313}.

\bibitem{Kachelriess:2006fi}
\bibinfo{author}{\bibfnamefont{M.}~\bibnamefont{Kachelrie{\ss}}}
  \bibnamefont{and}
  \bibinfo{author}{\bibfnamefont{R.}~\bibnamefont{Tom{\`a}s}},
  \bibinfo{journal}{Phys. Rev.} \textbf{\bibinfo{volume}{D74}},
  \bibinfo{pages}{063009} (\bibinfo{year}{2006}), \eprint{astro-ph/0606406}.

\bibitem{Rodejohann:2006qq}
\bibinfo{author}{\bibfnamefont{W.}~\bibnamefont{Rodejohann}},
\bibinfo{journal}{J. Cosmol. Astropart. Phys.}
\textbf{\bibinfo{volume}{01}}, \bibinfo{pages}{029}
(\bibinfo{year}{2007}), \eprint{hep-ph/0612047}.

\bibitem{Ginzburg:1990sk}
\bibinfo{author}{\bibfnamefont{V.~S.} \bibnamefont{Berezinsky}},
  \bibinfo{author}{\bibfnamefont{S.~V.} \bibnamefont{Bulanov}},
  \bibinfo{author}{\bibfnamefont{V.~A.} \bibnamefont{Dogiel}},
  \bibinfo{author}{\bibfnamefont{V.~L.} \bibnamefont{Ginzburg}},
  \bibnamefont{and} \bibinfo{author}{\bibfnamefont{V.~S.}
  \bibnamefont{Ptuskin}}, {\it Astrophysics of Cosmic Rays},
  \bibinfo{note}{Amsterdam, The Netherlands: North-Holland (1990) 534 p}.

\bibitem{Rachen:1998fd}
\bibinfo{author}{\bibfnamefont{J.~P.} \bibnamefont{Rachen}} \bibnamefont{and}
  \bibinfo{author}{\bibfnamefont{P.}~\bibnamefont{Meszaros}},
  \bibinfo{journal}{Phys. Rev.} \textbf{\bibinfo{volume}{D58}},
  \bibinfo{pages}{123005} (\bibinfo{year}{1998}), \eprint{astro-ph/9802280}.

\bibitem{Kashti:2005qa}
\bibinfo{author}{\bibfnamefont{T.}~\bibnamefont{Kashti}} \bibnamefont{and}
  \bibinfo{author}{\bibfnamefont{E.}~\bibnamefont{Waxman}},
  \bibinfo{journal}{Phys. Rev. Lett.} \textbf{\bibinfo{volume}{95}},
  \bibinfo{pages}{181101} (\bibinfo{year}{2005}), \eprint{astro-ph/0507599}.

\bibitem{Anchordoqui:2003vc}
\bibinfo{author}{\bibfnamefont{L.~A.} \bibnamefont{Anchordoqui}},
  \bibinfo{author}{\bibfnamefont{H.}~\bibnamefont{Goldberg}},
  \bibinfo{author}{\bibfnamefont{F.}~\bibnamefont{Halzen}}, \bibnamefont{and}
  \bibinfo{author}{\bibfnamefont{T.~J.} \bibnamefont{Weiler}},
  \bibinfo{journal}{Phys. Lett.} \textbf{\bibinfo{volume}{B593}},
  \bibinfo{pages}{42} (\bibinfo{year}{2004}), \eprint{astro-ph/0311002}.

\bibitem{Hooper:2004xr}
\bibinfo{author}{\bibfnamefont{D.}~\bibnamefont{Hooper}},
  \bibinfo{author}{\bibfnamefont{D.}~\bibnamefont{Morgan}}, \bibnamefont{and}
  \bibinfo{author}{\bibfnamefont{E.}~\bibnamefont{Winstanley}},
  \bibinfo{journal}{Phys. Lett.} \textbf{\bibinfo{volume}{B609}},
  \bibinfo{pages}{206} (\bibinfo{year}{2005}), \eprint{hep-ph/0410094}.

\bibitem{Winter:2006ce}
\bibinfo{author}{\bibfnamefont{W.}~\bibnamefont{Winter}},
  \bibinfo{journal}{Phys. Rev.} \textbf{\bibinfo{volume}{D74}},
  \bibinfo{pages}{033015} (\bibinfo{year}{2006}), \eprint{hep-ph/0604191}.

\bibitem{Acker:1991ej}
\bibinfo{author}{\bibfnamefont{A.}~\bibnamefont{Acker}},
  \bibinfo{author}{\bibfnamefont{S.}~\bibnamefont{Pakvasa}}, \bibnamefont{and}
  \bibinfo{author}{\bibfnamefont{J.~T.} \bibnamefont{Pantaleone}},
  \bibinfo{journal}{Phys. Rev.} \textbf{\bibinfo{volume}{D45}},
  \bibinfo{pages}{R1} (\bibinfo{year}{1992}).

\bibitem{Beacom:2002vi}
\bibinfo{author}{\bibfnamefont{J.~F.} \bibnamefont{Beacom}},
  \bibinfo{author}{\bibfnamefont{N.~F.} \bibnamefont{Bell}},
  \bibinfo{author}{\bibfnamefont{D.}~\bibnamefont{Hooper}},
  \bibinfo{author}{\bibfnamefont{S.}~\bibnamefont{Pakvasa}}, \bibnamefont{and}
  \bibinfo{author}{\bibfnamefont{T.~J.} \bibnamefont{Weiler}},
  \bibinfo{journal}{Phys. Rev. Lett.} \textbf{\bibinfo{volume}{90}},
  \bibinfo{pages}{181301} (\bibinfo{year}{2003}), \eprint{hep-ph/0211305}.

\bibitem{Hooper:2005jp}
\bibinfo{author}{\bibfnamefont{D.}~\bibnamefont{Hooper}},
  \bibinfo{author}{\bibfnamefont{D.}~\bibnamefont{Morgan}}, \bibnamefont{and}
  \bibinfo{author}{\bibfnamefont{E.}~\bibnamefont{Winstanley}},
  \bibinfo{journal}{Phys. Rev.} \textbf{\bibinfo{volume}{D72}},
  \bibinfo{pages}{065009} (\bibinfo{year}{2005}), \eprint{hep-ph/0506091}.

\bibitem{Gonzalez-Garcia:2005xw}
\bibinfo{author}{\bibfnamefont{M.~C.} \bibnamefont{Gonzalez-Garcia}},
  \bibinfo{author}{\bibfnamefont{F.}~\bibnamefont{Halzen}}, \bibnamefont{and}
  \bibinfo{author}{\bibfnamefont{M.}~\bibnamefont{Maltoni}},
  \bibinfo{journal}{Phys. Rev.} \textbf{\bibinfo{volume}{D71}},
  \bibinfo{pages}{093010} (\bibinfo{year}{2005}), \eprint{hep-ph/0502223}.

\bibitem{Anchordoqui:2005gj}
\bibinfo{author}{\bibfnamefont{L.~A.} \bibnamefont{Anchordoqui}} \emph{et~al.},
  \bibinfo{journal}{Phys. Rev.} \textbf{\bibinfo{volume}{D72}},
  \bibinfo{pages}{065019} (\bibinfo{year}{2005}), \eprint{hep-ph/0506168}.

\bibitem{Kobayashi:2000md}
\bibinfo{author}{\bibfnamefont{M.}~\bibnamefont{Kobayashi}} \bibnamefont{and}
  \bibinfo{author}{\bibfnamefont{C.~S.} \bibnamefont{Lim}},
  \bibinfo{journal}{Phys. Rev.} \textbf{\bibinfo{volume}{D64}},
  \bibinfo{pages}{013003} (\bibinfo{year}{2001}), \eprint{hep-ph/0012266}.

\bibitem{Beacom:2003eu}
\bibinfo{author}{\bibfnamefont{J.~F.} \bibnamefont{Beacom}} \emph{et~al.},
  \bibinfo{journal}{Phys. Rev. Lett.} \textbf{\bibinfo{volume}{92}},
  \bibinfo{pages}{011101} (\bibinfo{year}{2004}), \eprint{hep-ph/0307151}.

\bibitem{Mena:2006eq}
\bibinfo{author}{\bibfnamefont{O.}~\bibnamefont{Mena}},
  \bibinfo{author}{\bibfnamefont{I.}~\bibnamefont{Mocioiu}}, \bibnamefont{and}
  \bibinfo{author}{\bibfnamefont{S.}~\bibnamefont{Razzaque}},
  \bibinfo{journal}{Phys. Rev.} \textbf{\bibinfo{volume}{D75}},
  \bibinfo{pages}{063003} (\bibinfo{year}{2007}), \eprint{astro-ph/0612325}.

\bibitem{Blennow:2005yk}
\bibinfo{author}{\bibfnamefont{M.}~\bibnamefont{Blennow}},
  \bibinfo{author}{\bibfnamefont{T.}~\bibnamefont{Ohlsson}}, \bibnamefont{and}
  \bibinfo{author}{\bibfnamefont{W.}~\bibnamefont{Winter}},
  \bibinfo{journal}{J. High Energy Phys.}
  \textbf{\bibinfo{volume}{06}}, \bibinfo{pages}{049}
  (\bibinfo{year}{2005}), \eprint{hep-ph/0502147}.

\bibitem{Xing:2006xd}
\bibinfo{author}{\bibfnamefont{Z.-z.} \bibnamefont{Xing}},
  \bibinfo{journal}{Phys. Rev.} \textbf{\bibinfo{volume}{D74}},
  \bibinfo{pages}{013009} (\bibinfo{year}{2006}), \eprint{hep-ph/0605219}.

\bibitem{Xing:2006uk}
\bibinfo{author}{\bibfnamefont{Z.-Z.} \bibnamefont{Xing}} \bibnamefont{and}
  \bibinfo{author}{\bibfnamefont{S.}~\bibnamefont{Zhou}},
  \bibinfo{journal}{Phys. Rev.} \textbf{\bibinfo{volume}{D74}},
  \bibinfo{pages}{013010} (\bibinfo{year}{2006}), \eprint{astro-ph/0603781}.

\bibitem{Strumia:2006db}
\bibinfo{author}{\bibfnamefont{A.}~\bibnamefont{Strumia}} \bibnamefont{and}
  \bibinfo{author}{\bibfnamefont{F.}~\bibnamefont{Vissani}}
  (\bibinfo{year}{2006}), \eprint{hep-ph/0606054}.

\bibitem{Fogli:2006jk}
\bibinfo{author}{\bibfnamefont{G.~L.} \bibnamefont{Fogli}},
  \bibinfo{author}{\bibfnamefont{E.}~\bibnamefont{Lisi}},
  \bibinfo{author}{\bibfnamefont{A.}~\bibnamefont{Mirizzi}},
  \bibinfo{author}{\bibfnamefont{D.}~\bibnamefont{Montanino}},
  \bibnamefont{and} \bibinfo{author}{\bibfnamefont{P.~D.}
  \bibnamefont{Serpico}}, \bibinfo{journal}{Phys. Rev.}
  \textbf{\bibinfo{volume}{D74}}, \bibinfo{pages}{093004}
  (\bibinfo{year}{2006}), \eprint{hep-ph/0608321}.

\bibitem{Fukuyama:1997ky}
\bibinfo{author}{\bibfnamefont{T.}~\bibnamefont{Fukuyama}} \bibnamefont{and}
  \bibinfo{author}{\bibfnamefont{H.}~\bibnamefont{Nishiura}}
  (\bibinfo{year}{1997}), \eprint{hep-ph/9702253}.

\bibitem{Mohapatra:1998ka}
\bibinfo{author}{\bibfnamefont{R.~N.} \bibnamefont{Mohapatra}}
\bibnamefont{and} \bibinfo{author}{\bibfnamefont{S.}~\bibnamefont{Nussinov}},
  \bibinfo{journal}{Phys. Rev.} \textbf{\bibinfo{volume}{D60}},
  \bibinfo{pages}{013002} (\bibinfo{year}{1999}), \eprint{hep-ph/9809415}.

\bibitem{Ma:2001mr}
\bibinfo{author}{\bibfnamefont{E.} \bibnamefont{Ma}} \bibnamefont{and}
  \bibinfo{author}{\bibfnamefont{M.}~\bibnamefont{Raidal}},
  \bibinfo{journal}{Phys. Rev. Lett.} \textbf{\bibinfo{volume}{87}},
  \bibinfo{pages}{011802} (\bibinfo{year}{2001}), \eprint{hep-ph/0102255}.

\bibitem{Lam:2001fb}
\bibinfo{author}{\bibfnamefont{C.~S.} \bibnamefont{Lam}},
  \bibinfo{journal}{Phys. Lett.} \textbf{\bibinfo{volume}{B507}},
  \bibinfo{pages}{214} (\bibinfo{year}{2001}), \eprint{hep-ph/0104116}.

\bibitem{Harrison:2002et}
\bibinfo{author}{\bibfnamefont{P.~F.} \bibnamefont{Harrison}} \bibnamefont{and}
  \bibinfo{author}{\bibfnamefont{W.~G.}~\bibnamefont{Scott}},
  \bibinfo{journal}{Phys. Lett.} \textbf{\bibinfo{volume}{B547}},
  \bibinfo{pages}{219} (\bibinfo{year}{2002}), \eprint{hep-ph/0210197}.

\bibitem{Vissani:2006pi}
\bibinfo{author}{\bibfnamefont{F.}~\bibnamefont{Vissani}}
  (\bibinfo{year}{2006}), \eprint{astro-ph/0609575}.

\bibitem{Pakvasa:1999nh}
\bibinfo{author}{\bibfnamefont{S.}~\bibnamefont{Pakvasa}}
  (\bibinfo{year}{1999}), \eprint{hep-ph/9905426}.

\bibitem{Lipari:1999vh}
\bibinfo{author}{\bibfnamefont{P.}~\bibnamefont{Lipari}} \bibnamefont{and}
  \bibinfo{author}{\bibfnamefont{M.}~\bibnamefont{Lusignoli}},
  \bibinfo{journal}{Phys. Rev.} \textbf{\bibinfo{volume}{D60}},
  \bibinfo{pages}{013003} (\bibinfo{year}{1999}), \eprint{hep-ph/9901350}.

\bibitem{Barger:1999bg}
\bibinfo{author}{\bibfnamefont{V.~D.} \bibnamefont{Barger}} \emph{et~al.},
  \bibinfo{journal}{Phys. Lett.} \textbf{\bibinfo{volume}{B462}},
  \bibinfo{pages}{109} (\bibinfo{year}{1999}), \eprint{hep-ph/9907421}.

\bibitem{Ashie:2004mr}
\bibinfo{author}{\bibfnamefont{Y.}~\bibnamefont{Ashie}} \emph{et~al.}
  (\bibinfo{collaboration}{Super-Kamiokande}), \bibinfo{journal}{Phys. Rev.
  Lett.} \textbf{\bibinfo{volume}{93}}, \bibinfo{pages}{101801}
  (\bibinfo{year}{2004}), \eprint{hep-ex/0404034}.

\bibitem{Araki:2004mb}
\bibinfo{author}{\bibfnamefont{T.}~\bibnamefont{Araki}} \emph{et~al.}
  (\bibinfo{collaboration}{KamLAND}), \bibinfo{journal}{Phys. Rev. Lett.}
  \textbf{\bibinfo{volume}{94}}, \bibinfo{pages}{081801}
  (\bibinfo{year}{2005}), \eprint{hep-ex/0406035}.

\bibitem{Beacom:2002cb}
\bibinfo{author}{\bibfnamefont{J.~F.} \bibnamefont{Beacom}} \bibnamefont{and}
  \bibinfo{author}{\bibfnamefont{N.~F.} \bibnamefont{Bell}},
  \bibinfo{journal}{Phys. Rev.} \textbf{\bibinfo{volume}{D65}},
  \bibinfo{pages}{113009} (\bibinfo{year}{2002}), \eprint{hep-ph/0204111}.

\bibitem{Frieman:1987as}
\bibinfo{author}{\bibfnamefont{J.~A.} \bibnamefont{Frieman}},
  \bibinfo{author}{\bibfnamefont{H.~E.} \bibnamefont{Haber}}, \bibnamefont{and}
  \bibinfo{author}{\bibfnamefont{K.}~\bibnamefont{Freese}},
  \bibinfo{journal}{Phys. Lett.} \textbf{\bibinfo{volume}{B200}},
  \bibinfo{pages}{115} (\bibinfo{year}{1988}).

\bibitem{Fogli:2004gy}
\bibinfo{author}{\bibfnamefont{G.~L.} \bibnamefont{Fogli}},
  \bibinfo{author}{\bibfnamefont{E.}~\bibnamefont{Lisi}},
  \bibinfo{author}{\bibfnamefont{A.}~\bibnamefont{Mirizzi}}, \bibnamefont{and}
  \bibinfo{author}{\bibfnamefont{D.}~\bibnamefont{Montanino}},
  \bibinfo{journal}{Phys. Rev.} \textbf{\bibinfo{volume}{D70}},
  \bibinfo{pages}{013001} (\bibinfo{year}{2004}), \eprint{hep-ph/0401227}.

\bibitem{Chikashige:1980qk}
\bibinfo{author}{\bibfnamefont{Y.} \bibnamefont{Chikashige}},
  \bibinfo{author}{\bibfnamefont{R.~N.}~\bibnamefont{Mohapatra}},
  \bibnamefont{and} 
  \bibinfo{author}{\bibfnamefont{R.~D.}~\bibnamefont{Peccei}},
  \bibinfo{journal}{Phys. Rev. Lett.} \textbf{\bibinfo{volume}{45}},
  \bibinfo{pages}{1926} (\bibinfo{year}{1980}).

\bibitem{Gelmini:1980re}
\bibinfo{author}{\bibfnamefont{G.~B.} \bibnamefont{Gelmini}} \bibnamefont{and}
  \bibinfo{author}{\bibfnamefont{M.}~\bibnamefont{Roncadelli}},
  \bibinfo{journal}{Phys. Lett.} \textbf{\bibinfo{volume}{B99}},
  \bibinfo{pages}{411} (\bibinfo{year}{1981}).

\bibitem{Pakvasa:1999ta}
\bibinfo{author}{\bibfnamefont{S.}~\bibnamefont{Pakvasa}},
  \bibinfo{journal}{AIP Conf. Proc.} \textbf{\bibinfo{volume}{542}},
  \bibinfo{pages}{99} (\bibinfo{year}{2000}), \eprint{hep-ph/0004077}.

\bibitem{Hannestad:2005ex}
\bibinfo{author}{\bibfnamefont{S.} \bibnamefont{Hannestad}} \bibnamefont{and}
  \bibinfo{author}{\bibfnamefont{G.}~\bibnamefont{Raffelt}},
  \bibinfo{journal}{Phys. Rev.} \textbf{\bibinfo{volume}{D72}},
  \bibinfo{pages}{103514} (\bibinfo{year}{2005}), \eprint{hep-ph/0509278}.

\bibitem{Bell:2005dr}
\bibinfo{author}{\bibfnamefont{N.~F.} \bibnamefont{Bell}},
  \bibinfo{author}{\bibfnamefont{E.}~\bibnamefont{Pierpaoli}},
  \bibnamefont{and} 
  \bibinfo{author}{\bibfnamefont{K.}~\bibnamefont{Sigurdson}},
  \bibinfo{journal}{Phys. Rev.} \textbf{\bibinfo{volume}{D73}},
  \bibinfo{pages}{063523} (\bibinfo{year}{2006}), \eprint{astro-ph/0511410}.

\bibitem{Lindner:2001fx}
\bibinfo{author}{\bibfnamefont{M.}~\bibnamefont{Lindner}},
  \bibinfo{author}{\bibfnamefont{T.}~\bibnamefont{Ohlsson}}, \bibnamefont{and}
  \bibinfo{author}{\bibfnamefont{W.}~\bibnamefont{Winter}},
  \bibinfo{journal}{Nucl. Phys.} \textbf{\bibinfo{volume}{B607}},
  \bibinfo{pages}{326} (\bibinfo{year}{2001}), \eprint{hep-ph/0103170}.

\bibitem{Beacom:2003zg}
\bibinfo{author}{\bibfnamefont{J.~F.} \bibnamefont{Beacom}},
  \bibinfo{author}{\bibfnamefont{N.~F.} \bibnamefont{Bell}},
  \bibinfo{author}{\bibfnamefont{D.}~\bibnamefont{Hooper}},
  \bibinfo{author}{\bibfnamefont{S.}~\bibnamefont{Pakvasa}}, \bibnamefont{and}
  \bibinfo{author}{\bibfnamefont{T.~J.} \bibnamefont{Weiler}},
  \bibinfo{journal}{Phys. Rev.} \textbf{\bibinfo{volume}{D69}},
  \bibinfo{pages}{017303} (\bibinfo{year}{2004}), \eprint{hep-ph/0309267}.

\bibitem{beacom2003}
\bibinfo{author}{\bibfnamefont{J.~F.} \bibnamefont{Beacom}} \emph{et~al.},
  \bibinfo{journal}{Phys. Rev. D} \textbf{\bibinfo{volume}{68}},
  \bibinfo{pages}{093005} (\bibinfo{year}{2003}), \eprint{hep-ph/0307025}.

\bibitem{Learned:1994wg}
\bibinfo{author}{\bibfnamefont{J.~G.} \bibnamefont{Learned}} \bibnamefont{and}
  \bibinfo{author}{\bibfnamefont{S.}~\bibnamefont{Pakvasa}},
  \bibinfo{journal}{Astropart. Phys.} \textbf{\bibinfo{volume}{3}},
  \bibinfo{pages}{267} (\bibinfo{year}{1995}), \eprint{hep-ph/9405296}.

\bibitem{Athar:2000rx}
\bibinfo{author}{\bibfnamefont{H.} \bibnamefont{Athar}},
  \bibinfo{author}{\bibfnamefont{G.}~\bibnamefont{Parente}}, \bibnamefont{and}
  \bibinfo{author}{\bibfnamefont{E.}~\bibnamefont{Zas}},
  \bibinfo{journal}{Phys. Rev. D} \textbf{\bibinfo{volume}{62}},
  \bibinfo{pages}{093010} (\bibinfo{year}{2000}), \eprint{hep-ph/0006123}.

\bibitem{Halzen:1998be}
\bibinfo{author}{\bibfnamefont{F.} \bibnamefont{Halzen}} \bibnamefont{and}
  \bibinfo{author}{\bibfnamefont{D.}~\bibnamefont{Saltzberg}},
  \bibinfo{journal}{Phys. Rev. Lett.} \textbf{\bibinfo{volume}{81}},
  \bibinfo{pages}{4305} (\bibinfo{year}{1998}), \eprint{hep-ph/9804354}.

\bibitem{Dutta:2000jv}
\bibinfo{author}{\bibfnamefont{S.~I.} \bibnamefont{Dutta}},
  \bibinfo{author}{\bibfnamefont{M.~H.}~\bibnamefont{Reno}}, \bibnamefont{and}
  \bibinfo{author}{\bibfnamefont{I.}~\bibnamefont{Sarcevic}},
  \bibinfo{journal}{Phys. Rev. D} \textbf{\bibinfo{volume}{62}},
  \bibinfo{pages}{123001} (\bibinfo{year}{2000}), \eprint{hep-ph/0005310}.


\bibitem{alvarez}
\bibinfo{author}{\bibfnamefont{J.}~\bibnamefont{Alvarez-Mu{\~n}iz}}
  \emph{et~al.}, \bibinfo{journal}{Phys. Rev. D} \textbf{\bibinfo{volume}{65}},
  \bibinfo{pages}{124015} (\bibinfo{year}{2002}), \eprint{hep-ph/0202081}.


\end{thebibliography}
\end{document}